\documentclass[twocolumn]{aastex63}

\newcommand\degree{^{\circ}}
\usepackage{wasysym}
\usepackage{colortbl}

\definecolor{mgm}{rgb}{0.0, 0.0, 1.0}

\definecolor{lwf}{rgb}{0.255, 0.0, 0.177}

\definecolor{trq}{rgb}{0.4, 0.8, 0.3}

\graphicspath{{./}{figures/}}

\begin{document}

\shorttitle{K2-138 Resonance Chain}
\shortauthors{MacDonald et al.}


\title{Confirming the 3:2 Resonance Chain of K2-138}

\author[0000-0002-5289-4582]{Mariah G. MacDonald}
\affiliation{Department of Astronomy \& Astrophysics, Center for Exoplanets and Habitable Worlds, The Pennsylvania State University, University Park, PA 16802, USA}
\affiliation{Department of Physics, The College of New Jersey, 2000 Pennington Road, Ewing, NJ 08628, USA}
\author[0000-0003-2372-1364]{Leonard Feil}
\author[0000-0002-8974-8095]{Tyler Quinn}
\affiliation{Department of Astronomy \& Astrophysics, Center for Exoplanets and Habitable Worlds, The Pennsylvania State University, University Park, PA 16802, USA}

\author[0000-0001-6009-8685]{David Rice}
\affiliation{Department of Physics \& Astronomy, University of Nevada, Las Vegas, Las Vegas, NV, 89154, USA}

\begin{abstract}
The study of orbital resonances allows for the constraint of planetary properties of compact systems. K2-138 is an early K-type star with six planets, five of which have been proposed to be in the longest chain of 3:2 mean motion resonances. To observe and potentially verify the resonant behavior of K2-138's planets, we run \textit{N}-body simulations using previously measured parameters. Through our analysis, we find that 99.2\% of our simulations result in a chain of 3:2 resonances, although only 11\% of them show a five-planet resonance chain. We find we are able to use resonances to constrain the orbital periods and masses of the planets. We explore the possibility of this system forming in situ and through disk migration, and investigate the potential compositions of each planet using a planet structure code. 
\end{abstract}



\section{Introduction} \label{sec:intro}
The Kepler and K2 missions allowed for the study of worlds other than our own and the systems they inhabit. Since the launch in 2009, we have confirmed more than 4,000 exoplanets, with several thousands of other candidates being investigated. This catalog of planets has enabled the expansion of various sub-fields of astronomy, including astrobiology, the study of atmospheres, and orbital dynamics and evolution.

Some of the systems we have discovered exhibit resonant behavior, including Kepler-223 \citep{Mills2017}, Kepler-80 \citep{MacDonald_2016}, and TRAPPIST-1 \citep{TRAPPISTRes}; mean motion resonance occurs when two or more planets repeatedly exchange angular momentum and energy as they orbit their host star, often seen as a repeated geometric configuration. We can predict a system's resonances by observing the orbital periods of the planets, as planets in or near mean motion resonance have period ratios that reduce to a ratio of small numbers. However, a period ratio near commensurability does not guarantee a resonance; we must study the system's dynamics and resonance angles to confirm resonance. We describe a two-body resonance by the libration, or oscillation, of the two-body angle:
\begin{equation}\label{twobodyres}
\Theta_{b,c} = j_1\lambda_b + j_2\lambda_c + j_3\omega_b + j_4\omega_c + j_5\Omega_b + j_6\Omega_c
\end{equation}

\noindent where $\lambda$ is the mean longitude, $\omega$ is the argument of periapsis, $\Omega$ is the longitude of the ascending node, planet b is interior to planet c, and the coefficients $j_i$ sum to zero.

A system can have more than two planets in resonance, either in a chain of librating two-body resonances, or in a three- or more body resonance. The three-body angle is the difference between the associated two-body resonance angles:

\begin{equation}\label{threebodyres}
    \phi = p\lambda_3 - (p + q)\lambda_2 + q\lambda_1
\end{equation}

\noindent where $\lambda_i$ is again the mean longitude and $p$ and $q$ are integers. It is easier, and therefore more common, to constrain the libration of three-body resonance angles than that of two-body angles since the two-body angles depend on the longitudes of periapsis which are challenging to constrain for low-eccentricity exoplanets.

By observing a potentially resonant system with high enough precision photometry or high enough cadence, we are able to measure the resonance angles and confirm resonance. However, such data do not yet exist for most systems, and therefore we must model the system across all planetary and orbital parameters that are consistent with the data; if all parameters lead to solutions with librating angles, we are able to confirm the resonance(s) of the system.

K2-138 is a relatively bright ($V = 12.21$) K-dwarf, hosting six confirmed planets\footnote{K2-138g was most recently confirmed by \citet{Hardegree2021}}. These inner five planets orbit their star fairly rapidly, with orbital periods ranging from 2.4 days to 12.8 days. In addition, the period ratios of adjacent planets suggest that the system could be locked in a five planet chain of 3:2 mean motion resonances, the longest 3:2 resonance chain known if confirmed. Both \citet{christiansen2018k2} and \citet{lopez2019exoplanet} suggest this chain, however, no study has yet performed an in-depth study of the orbital dynamics of K2-138 to confirm such a chain.

Here, we perform such a study with the aim of confirming the resonance chain and constraining the system's formation and dynamical evolution. In Section~\ref{sec:methods}, we discuss our \textit{N}-body simulations. We then present our results in Section \ref{sec:Results}. We use the system's resonances to constrain the planetary masses and orbital periods and discuss various pathways for forming the chain in Section~\ref{sec:discussion}. We also discuss the planetary compositions before summarizing and concluding our work in Section~\ref{sec:conclusion}.


\section{Methods}\label{sec:methods}
To observe the long term behavior of K2-138 and to constrain the dynamics of the system, we run \textit{N}-body simulations using REBOUND \citep{rebound}. We model the system with a stellar mass of $M_{\star}=0.93M_{\odot}$ \citep{christiansen2018k2}, and use the orbital parameters as constrained by \citet{lopez2019exoplanet}. We do not model the outer-most planet K2-138g since it is most likely dynamically decoupled with an orbital period of 42 days \citep[][]{Lopez2019,Hardegree2021}. For each simulation, we draw planetary masses, inclinations, and orbital periods from normal distributions centered on the nominal values from \citet{lopez2019exoplanet} with standard deviations equal to the uncertainties. We use the WHFast integrator \citep{rein2015} and integrate the system for 8Myr with a timestep of 5\% of the innermost planet's period.  We summarize the initial conditions of our simulations in Table~\ref{tab:sim_parameters}.

\begin{deluxetable*}{lccccc}
\renewcommand{\arraystretch}{0.75}
\tablecolumns{6}
\tablewidth{0pt}
\tabletypesize{\small}
\tablecaption{ Planetary Properties of K2-138 \label{tab:sim_parameters}}
\tablehead{
\colhead{} &
\colhead{K2-138 b} &
\colhead{K2-138 c} &
\colhead{K2-138 d} &
\colhead{K2-138 e} &
\colhead{K2-138 f}
}
\startdata
    $P$ [d] & $2.3531 \pm 0.0002$ & $3.5600\pm0.0001$ & $5.4048 \pm 0.0002$ & $8.2615\pm0.0002$ & $12.7576\pm0.0005$ \\
    $\textrm{t}_0$ [d] & $73.3168\pm0.0009$ & $40.32182\pm0.0009$ & $43.1599\pm0.0009$ & $40.6456^{+0.0009}_{-0.0008}$ & $38.7033^{+0.0009}_{-0.0009}$ \\
    $i$ [$\degree$] & $87.2^{+1.2}_{-1.0}$ & $88.1 \pm 0.7$ & $89.0 \pm 0.6$ & $88.6 \pm 0.3$ & $88.8 \pm 0.2$ \\
    $M_p$ [$M_{\oplus}$] &  $3.1 \pm 1.1$ & $6.3^{+1.1}_{-1.2}$ & $7.9^{+1.4}_{-1.3}$ & $13.0 \pm 2.0$ & $1.6^{+2.1}_{-1.2}$ \\
\enddata
\tablecomments{Parameters for the \textit{N}-body simulations of K2-138. Here, $P$ is the orbital period, $t_0$ represents the transit epoch (BJD$-2457700$), $i$ is the sky-plane inclination, and $M_p$ is the planetary mass. We use the values published by \citet{lopez2019exoplanet} for all parameters, and a stellar mass of $0.9310^{+0.0700}_{-0.0640}$ \citep{christiansen2018k2}.}
\end{deluxetable*}


\section{Results}\label{sec:Results}
We run 3000 \textit{N}-body simulations for 8Myr and analyze the results of each to confirm a resonance chain. We look for libration of all of the two-body angles, all of the three-body angles, or any combination of angles that leads to all planets participating in the chain. We summarize the dynamical results of our simulations in Table~\ref{tab:res_results}. 

\begin{figure*}[ht!]
    \centering
    \includegraphics[width=0.45\textwidth]{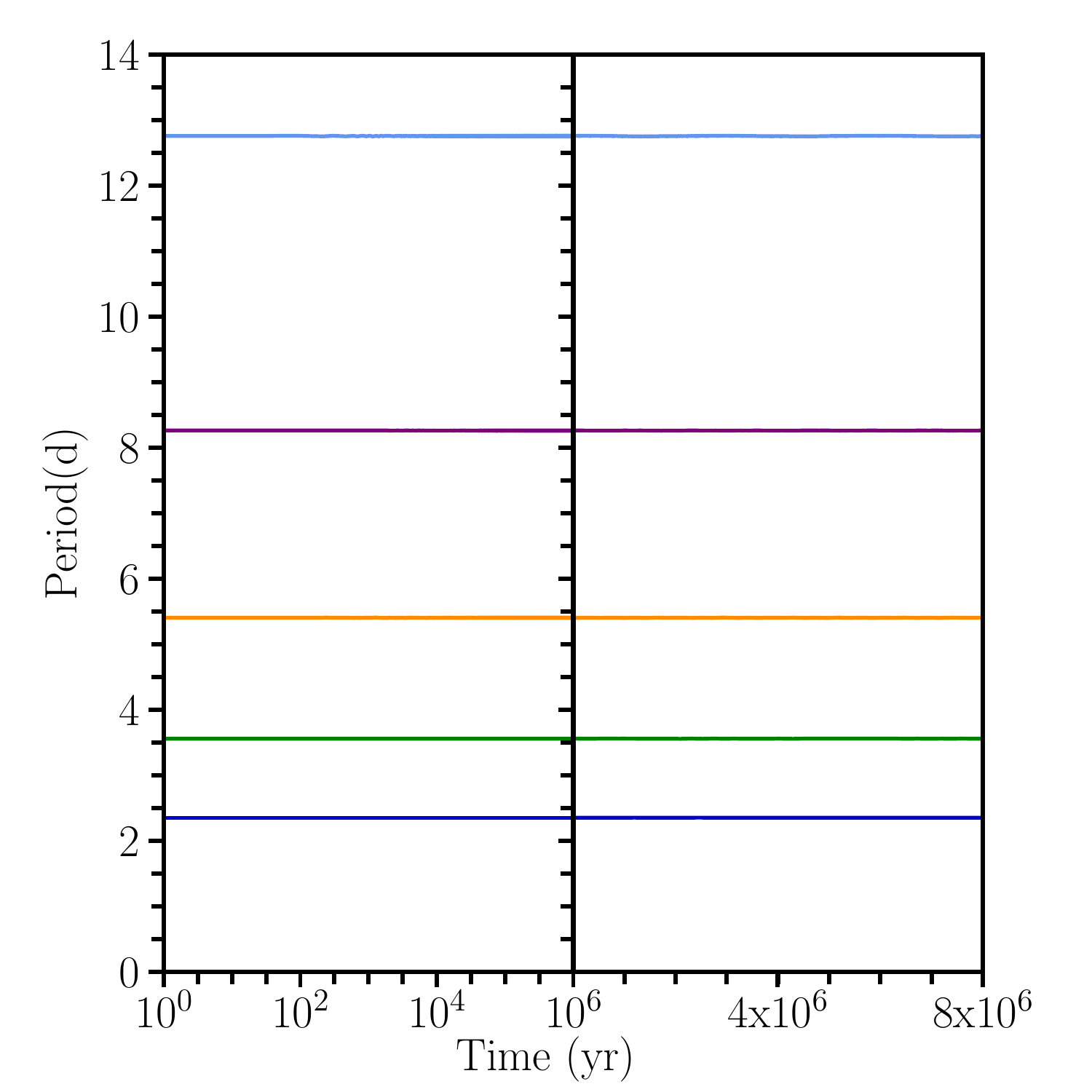}
    \includegraphics[width=0.45\textwidth]{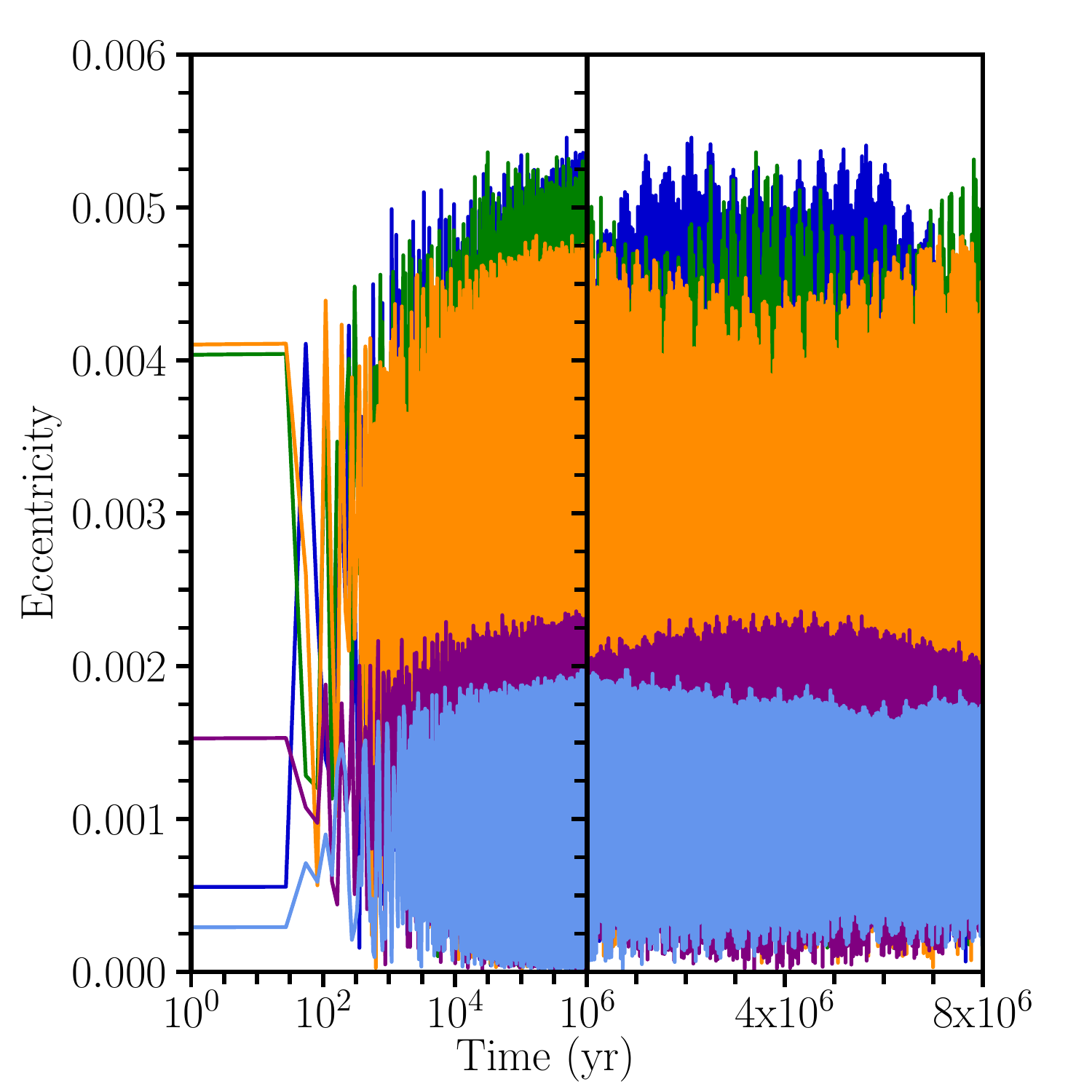}
    \includegraphics[width=0.45\textwidth]{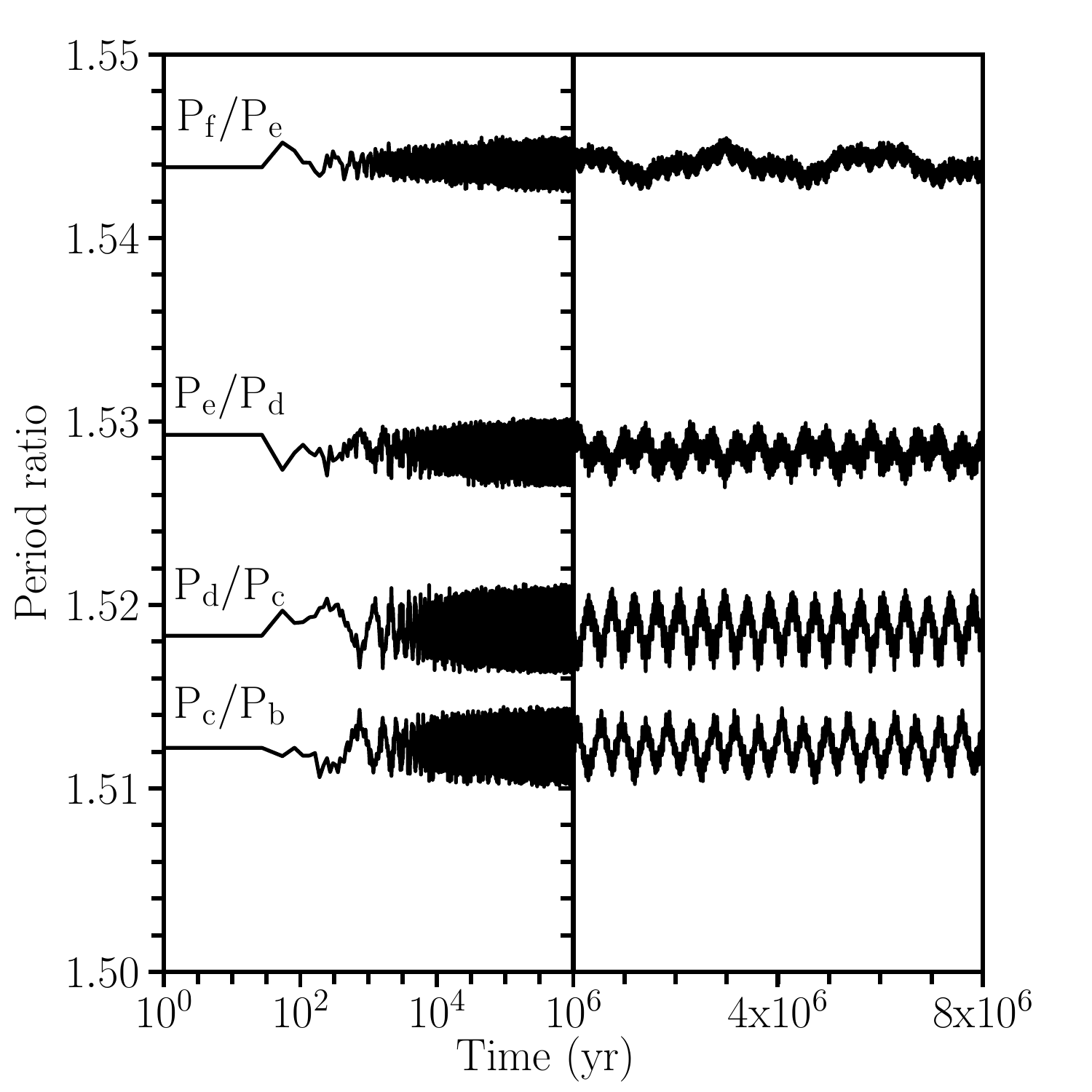}
    \includegraphics[width=0.45\textwidth]{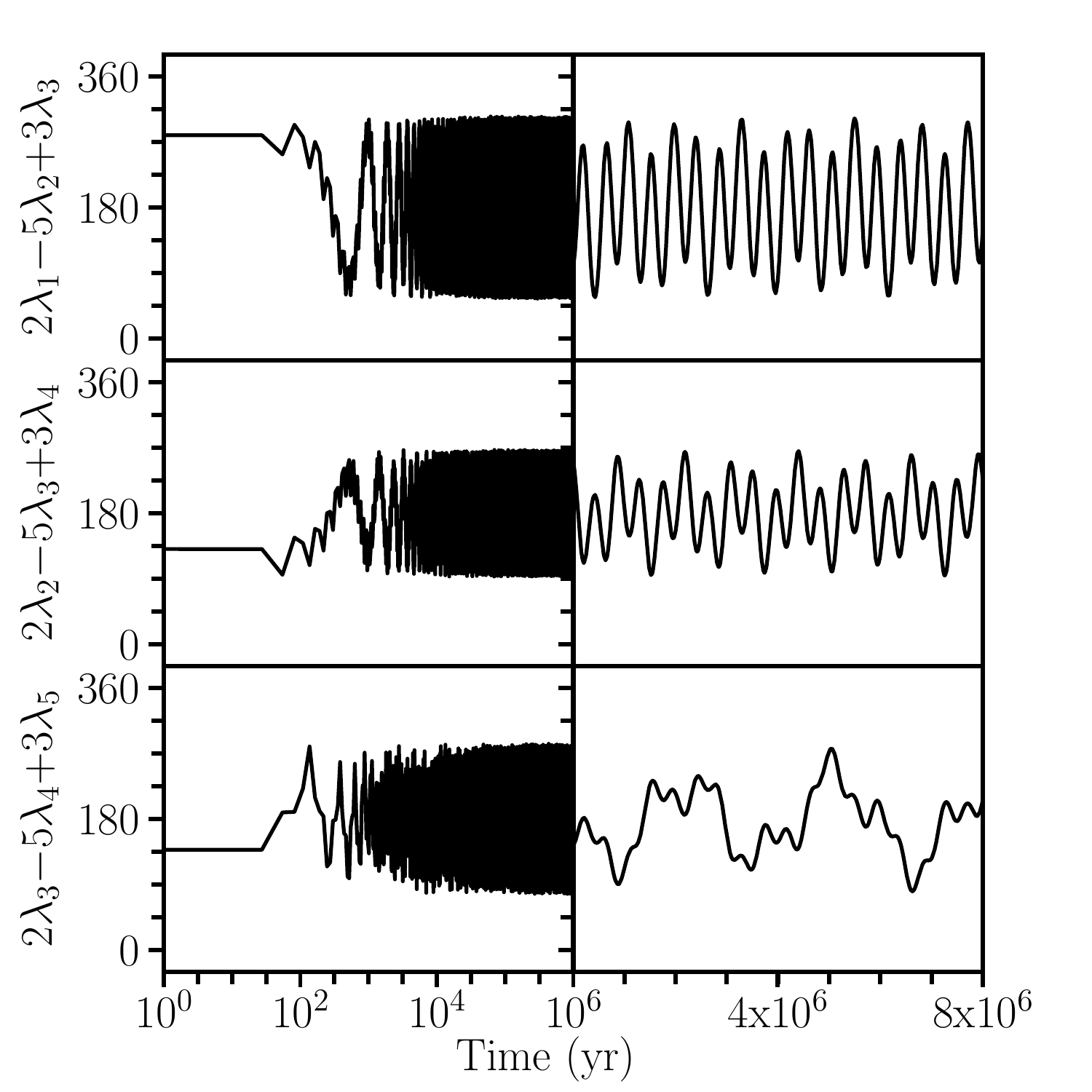}
    \caption{Evolution of the orbital periods, eccentricities, period ratios, and three-body resonance angles in an example simulation of K2-138. Here, we define resonance as the libration of a resonance angle, which occurs if the angle oscillates between two values. We say that the system contains a resonance chain if three or more planets are in resonance with one another. }
    \label{fig:k2_res}
\end{figure*}

We find that 99.2\% of our simulations result in a chain of 3:2 resonances, although only 11.0\% of our simulations result in a a five-planet resonance chain; in 87.1\% of the simulations, planet f is dynamically decoupled from the other planets. Overall, we find that 68.5\% of the simulations result in a four-planet resonance chain, and 19.6\% of the simulations result in only a three-planet resonance chain. Of the 0.8\% of our simulations where the planets are not interacting via a resonance chain, 76\% of the simulations result in no three-body angles librating and only the two-body angle between K2-138b and K2-138c and the angle between K2-138d and K2-138e librating; the remaining 24\% result in libration of only the two-body angle between K2-138d and K2-138e. We show an example of a fully librating five-planet resonance chain in Figure~\ref{fig:k2_res}.


\begin{deluxetable}{lllc}
\tablecolumns{4}
\tablewidth{0pt}
\tabletypesize{\footnotesize}
\tablecaption{Resonance Angles of K2-138 \label{tab:res_results}}
\tablehead{
\colhead{Angle} &
\colhead{\% Res} &
\colhead{Center [$\degree$]} &
\colhead{Amplitude [$\degree$]}
}
\startdata
$\phi_1 = 2\lambda_b-5\lambda_c+3\lambda_d$ & 28.56	&	179.76$^{+	5.05	}_{	4.45	} $ &	57.39$^{+	19.15	}_{	18.29	}$ \\
$\phi_2 = 2\lambda_c-5\lambda_d+3\lambda_e$ &  72.60	&	179.96$^{+	3.75	}_{	3.94	}$ &	49.31$^{+	22.89	}_{	16.09	}$ \\
$\phi_3 = 2\lambda_d-5\lambda_e+3\lambda_f$ &  11.46	&	180.21$^{+	5.94	}_{	5.15	}$ &	57.94$^{+	26.99	}_{	27.33	}$ \\
$\Theta_{c-b} = 3\lambda_c-2\lambda_b-\Omega_c$ & 64.35	&	-0.03$^{+	6.35	}_{	5.94	} $ &	61.78$^{+	9.93	}_{	13.30	}$ \\
$\Theta_{d-c} = 3\lambda_d-2\lambda_c-\Omega_d$ & 95.56	&	0.04$^{+	5.61	}_{	5.59	} $ &	55.18$^{+	11.05	}_{	11.90	}$ \\
$\Theta_{e-d} = 3\lambda_e-2\lambda_d-\Omega_e$ & 98.91	&	0.09$^{+	4.29	}_{	4.05	} $ &	39.95$^{+	14.66	}_{	11.07	}$ \\
$\Theta_{f-e} = 3\lambda_f-2\lambda_e-\Omega_f$ & 8.09	&	0.73$^{+	4.27	}_{	4.28	} $ &	57.83$^{+	16.45	}_{	15.60	}$ \\
 & 1.81	&	176.80$^{+	34.81	}_{	29.46	} $ &	88.06$^{+	7.50	}_{	5.96	}$
\enddata
\tablecomments{Resulting three-body and two-body angles from the REBOUND \textit{N}-body simulations. We find that 99.2\% of our simulations result in a resonance chain of 3:2 resonances, although only 11.0\% result in a five-planet resonance chain. Planet f is dynamically decoupled from the other planets in most of the simulations (87.1\%), while planet b is dynamically decoupled in 21.4\% of the simulations, planet c is dynamically decoupled in 0.2\% of the simulations, and planet e is dynamically decoupled in 0.17\% of the simulations.}
\end{deluxetable}

Given these results, we are able to confirm that the planets of K2-138 are indeed in a chain of 3:2 mean motion resonances, but we are not able to confirm a five-planet chain. The middle three planets --- K2-138c, K2-138d, and K2-138e --- are in resonance with one-another, but K2-138b and K2-138f do not need to be in resonance with other planets for the system to be stable or to be consistent with the data.


\section{Discussion}\label{sec:discussion}
Our motivation behind developing this method stems from the need to confirm more planetary candidates and diversify the planetary catalogue. In compact systems, resonant behavior could be a common means for maintaining stability \citep[e.g., ][]{tamayo2017}. Finding these configurations allows further constraints on the mass of candidates, which we can use to confirm their planetary identity.

In the following subsections, we constrain the planetary masses and orbital properties using the resonances, explore why K2-138f is not part of the chain, and discuss both the resonance chain's formation and the composition of K2-138's planets.




\subsection{Using Resonance to Constrain Mass}\label{sec:constrain}

Since we run our simulations to explore a large range of mass and orbital properties, we analyze our results to see which planetary parameters lead to the libration of the resonance angles. We first compare the distributions of planetary mass, eccentricity, and orbital period for simulations where each angle is librating to distributions of the same parameter from simulations where each angle is not librating using a two-sample Kolmogorov–Smirnov test. In this test, the null hypothesis is that the two samples (parameter from librating simulations and parameter from circulating simulations) are drawn from the same population, and the resulting \textit{p}-value is the probability that the null hypothesis is correct. Therefore, a small \textit{p}-value ($p<0.05$) allows us to reject this hypothesis and suggests that the two distributions are statistically distinct. 

\begin{figure*}[ht!]
    \centering
    \includegraphics[width=0.41\textwidth]{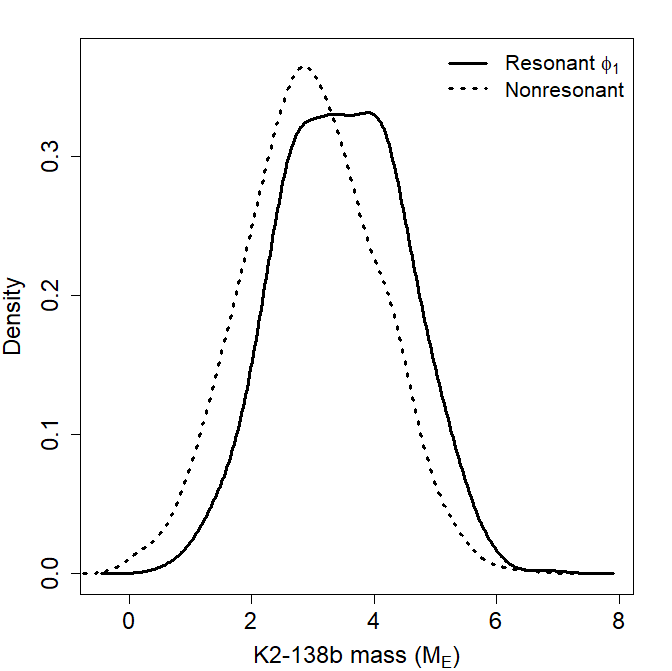}
    \includegraphics[width=0.41\textwidth]{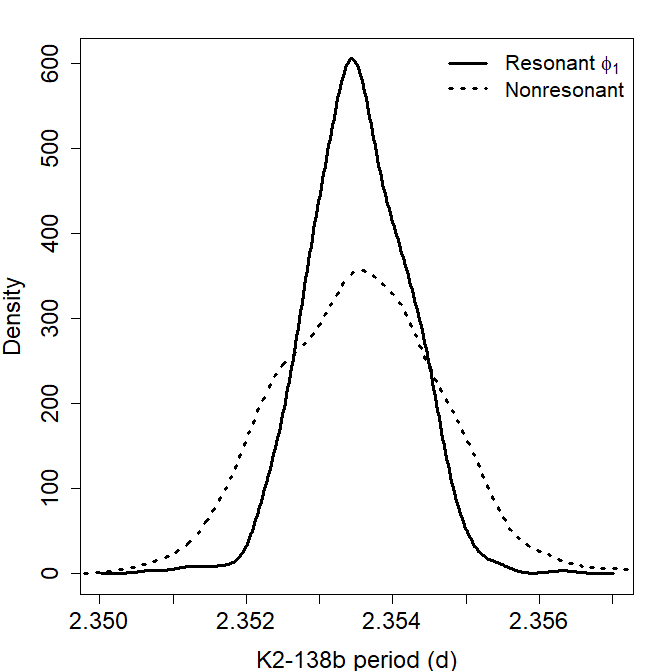}
    \includegraphics[width=0.41\textwidth]{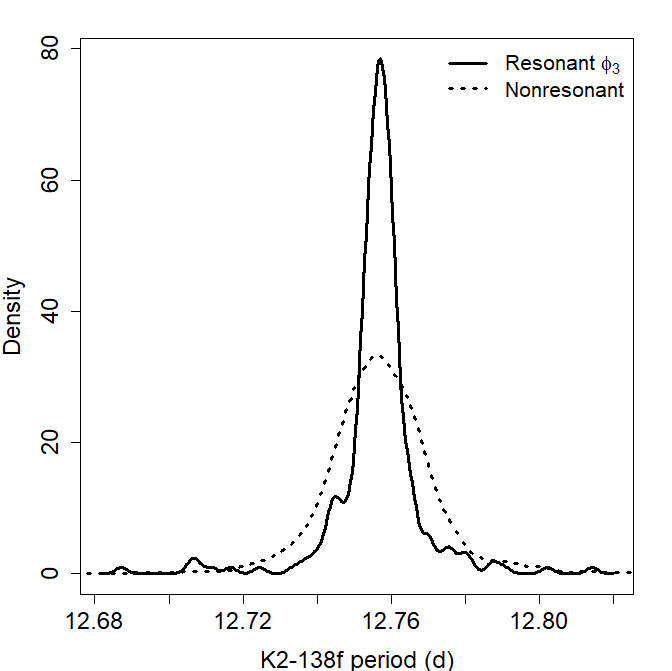}
    \includegraphics[width=0.41\textwidth]{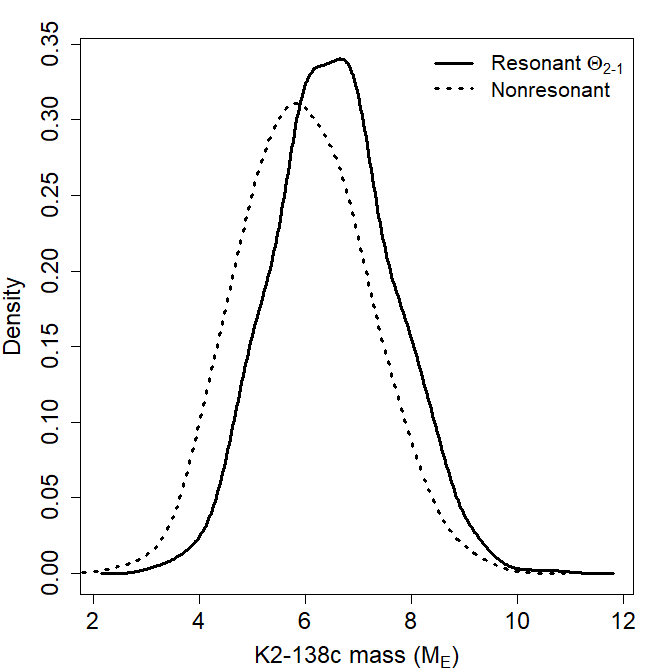}
    \includegraphics[width=0.41\textwidth]{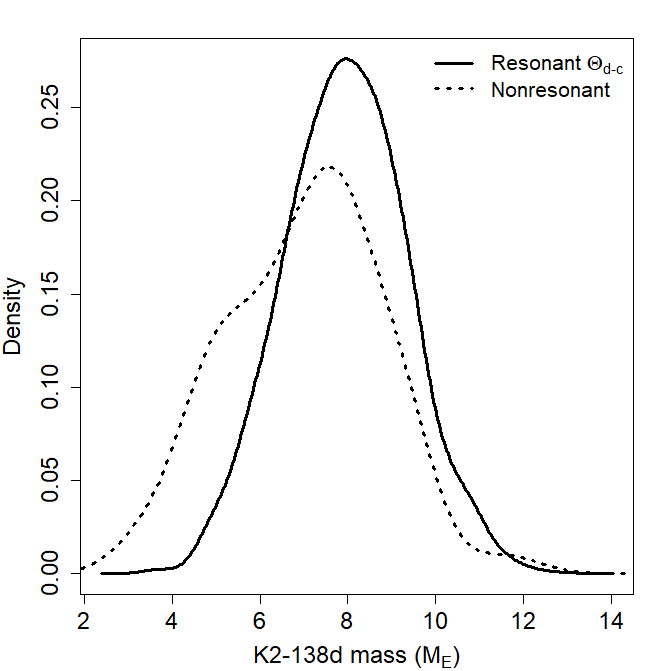}
    \includegraphics[width=0.41\textwidth]{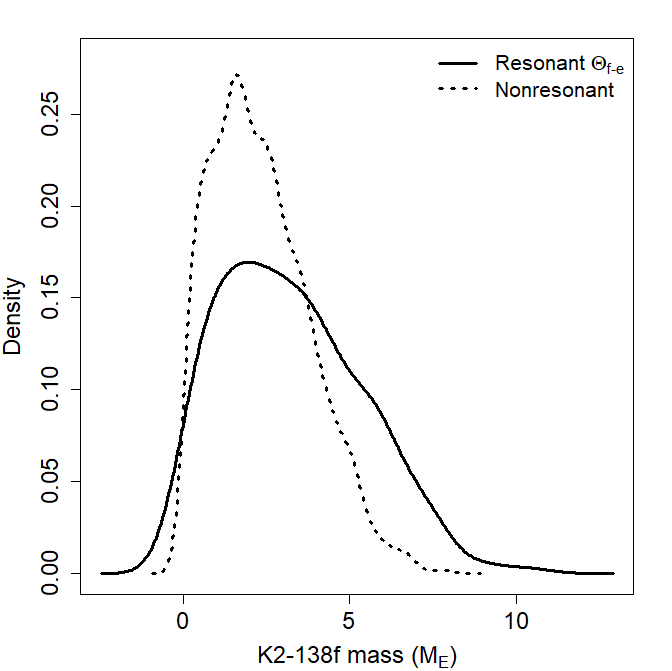}
    \caption{Kernel Density Estimations of various planetary masses and orbital periods, separated by whether a resonance angle was librating. For K-S \textit{p}-values and parameter estimates from this analysis, see Table~\ref{tab:KS}.}
    \label{fig: kdes}
\end{figure*}

We find that the libration of the three-body angle between the inner three planets depends on the eccentricities and orbital periods of the three planets but only depends on the mass of K2-138b. Similarly, the libration of the three-body angle between the outer three planets depends on the eccentricities of K2-138d and f, the orbital periods of the three planets, and the mass of K2-138f. The three-body angle between the middle planets and the four two-body angles also depend on some mixture of masses, eccentricities, and orbital periods. We show Kernel Density Estimations of the distributions of some of the more interesting dependencies in Figure~\ref{fig: kdes} and report the \textit{p}-values resulting from our K-S tests involving mass and orbital period in Table~\ref{tab:KS}.

For each angle and planet property pair with a small K-S \textit{p}-value, we report the median and 68.3\% confidence interval of the parameter in the simulations where the angle is librating in Table~\ref{tab:KS}. To further quantify which planet properties are statistically distinct between our simulations with librating resonance angles and the initial distributions, we perform a T-test on each parameter with a small K-S \textit{p}-value. The T-test null hypothesis states that there is no statistical difference between two groups, and a small \textit{p}-value indicates that an observed difference is not due to chance. We find that the masses of K2-138b, K2-138c, and K2-138f are statistically distinct between our simulations with librating resonance angles and the estimates from \citet{Lopez2019}. Our results suggest that these three planets are slightly more massive than the radial velocity data can constrain, with masses of $3.46	^{+	1.07}_{-1.01}$, 6.53$^{+1.17}_{-1.13}$, and 2.65$^{+1.90}_{-1.62}~M_{\oplus}$, respectively. If we constrain the mass of K2-138f using the two-body angle between it and K2-138e, we would recover an even larger mass of 3.01$^{+2.58}_{-1.98}~M_{\oplus}$, but the libration of this angle also depends on the orbital period of K2-138e.

The periods of all five planets shift slightly within the beginning of our simulations, even when they do not lock into resonance, and this shift results in a distribution of final orbital periods that is statistically distinct from those measured by \citet{Lopez2019} for all planets except K2-138f. Because of this, we compare the orbital periods of the planets between simulations with librating resonance angles and simulations without libration via the T-test. We recover large \textit{p}-values for all orbital periods except K2-138e, suggesting that our small K-S p-values are due to chance. For the two-body angle between K2-138e and K2-138f to librate, K2-138e requires a slightly larger period of $8.262\pm0.002$ days than that measured by \citet{Lopez2019} ($8.2615\pm0.0002$ days).


The differences in the masses of K2-138b, K2-138c, and K2-138f and in the orbital period of K2-138e could explain why we do not recover a five-planet resonance chain in all of our simulations. We discuss potential reasons why K2-138f is dynamically decoupled in the majority of our simulations below in Section~\ref{sec:whyf}.


\begin{deluxetable}{cccc}
\tablecolumns{4}
\tabletypesize{\footnotesize}
\tablecaption{\textit{p}-values and constraints from K-S Tests\label{tab:KS}}
\tablehead{
\colhead{Parameter} &
\colhead{Angle} &
\colhead{K-S $p$-value} &
\colhead{Estimate} }
\startdata
$	M_b	$ & 	$\phi_1	$ & 	1.00E-16	&	*3.46	$^{+	1.07	}_{-	1.01	}$ \\
$	M_b	$ & 	$\Theta_{c-b}	$ & 	1.1837E-66$^{\dagger}$	&	*3.36	$^{+	1.03	}_{-	1.02	}$ \\
$	M_c	$ & 	$\Theta_{c-b}	$ & 	1.36E-32$^{\dagger}$	&	*6.53	$^{+	1.17	}_{-	1.13	}$ \\
$	M_c	$ & 	$\Theta_{d-c}	$ & 	3.00E-04	&	6.33	$^{+	1.23	}_{-	1.17	}$ \\
$	M_d	$ & 	$\phi_2	$ & 	3.48E-02	&	7.93	$^{+	1.40	}_{-	1.37	}$ \\
$	M_d	$ & 	$\Theta_{d-c}	$ & 	2.65E-05	&	7.96	$^{+	1.37	}_{-	1.40	}$ \\
$	M_e	$ & 	$\phi_2	$ & 	1.80E-03	&	13.05	$^{+	1.96	}_{-	2.00	}$ \\
$	M_f	$ & 	$\phi_3	$ & 	1.19E-04	&	*2.65	$^{+	1.90	}_{-	1.62	}$ \\
$	M_f	$ & 	$\Theta_{f-e}	$ & 	4.84E-09	&	*3.01	$^{+	2.58	}_{-	1.98	}$ \\
$	P_b	$ & 	$\phi_1	$ & 	9.14E-09	&	*2.3535	$^{+	0.0007	}_{-	0.0006	}$ \\
$	P_c	$ & 	$\phi_1	$ & 	1.64E-03	&	*3.5592	$^{+	0.0010	}_{-	0.0009	}$ \\
$	P_c	$ & 	$\phi_2	$ & 	1.30E-04	&	*3.559	$\pm$	0.001	\\		
$	P_c	$ & 	$\Theta_{c-b}	$ & 	9.09E-03	&				\\		
$	P_d	$ & 	$\phi_1	$ & 	5.47E-03	&	*5.405	$\pm$	0.002	\\		
$	P_d	$ & 	$\phi_2	$ & 	5.96E-03	&				\\		
$	P_d	$ & 	$\phi_3	$ & 	9.24E-02	&				\\		
$	P_e	$ & 	$\phi_2	$ & 	3.37E-05	&	*8.262	$\pm$	0.002	\\		
$	P_e	$ & 	$\phi_3	$ & 	2.19E-03	&	*8.262	$^{+	0.001	}_{-	0.002	}$ \\
$	P_e	$ & 	$\Theta_{f-e}	$ & 	1.19E-02	&	*8.262	$\pm$	0.002	\\		
$	P_f	$ & 	$\phi_3	$ & 	4.53E-07	&	12.757	$\pm$	0.005	\\		
$	P_f	$ & 	$\Theta_{f-e}	$ & 	2.48E-08	&						
\enddata
\tablecomments{Resulting $p$-values from our two-sample Kolmogorov–Smirnov test, where one sample is the distribution of values from simulations where the resonance angle (Angle) is librating and the other sample is the values from circulating simulations. Here, the null hypothesis states that these two samples are drawn from the same population. We include the median and 68.3\% confidence intervals of each parameter from the simulations where the resonance angle is librating (Estimate).  Empty cells indicate the same values as the previous row. *: statistically distinct from \citet{Lopez2019} estimates.}
\end{deluxetable}

Ideally, we would constrain the masses and orbital parameters of the planets by fitting the TTVs of the planets or by photodynamically fitting the system. The TTVs, although estimated to be on the order of 2--5min \citep{christiansen2018k2,lopez2019exoplanet}, have so-far been illusive, indicating that the cadence or precision of the data is insufficient to tease out the perturbations. We summarize our photodynamic fitting efforts in the Appendix.




\subsection{Resonance of K2-138f \label{sec:whyf}}

We hypothesize three reasons why K2-138f is not part of the resonance chain in the majority of our simulations: 
\begin{enumerate}
    \item The planet is in resonance, but its orbital parameters, mass, and/or the masses or orbits of the other planets fall in the part of parameter space that is narrower than the data currently constrain (H1)
    \item The planet was once in resonance but this resonance has since been broken or disrupted (H2)
    \item The planet is not and was never in resonance (H3)
\end{enumerate}
We explore each of these hypotheses below.

\subsubsection{H1: Insufficient data}
The libration of a resonance angle depends on the masses and orbits of the participating planets, but also on the libration of other resonance angles in the system. For example, if three planets are near a chain of two MMRs, the libration of the two-body angle between the inner two planets will affect the libration of the two-body angle between the outer two planets. Because of this affect, an angle's libration might depend on other planets in the system.

To test this hypothesis, we run four additional suites of 500 \textit{N}-body simulations each to explore the likelihood of $\Theta_{f-e}$ and $\phi_3$ librating: one suite where we alter K2-138f's mass, one suite where we alter its orbital period, one suite where we increase the period of K2-138e, and one suite where we model the outermost planet K2-138g. We describe the results of each below.

\textit{Mass of f:} We increase and narrow the mass range of K2-138f to $2.65^{+1.90}_{-1.62}~M_{\oplus}$ (compared to $1.63^{+2.12}_{-1.98}~M_{\oplus}$). This increase in mass nearly triples the likelihood of both $\Theta_{f-e}$ and $\phi_3$ librating, as the angles librate in 28.9\% and 33.3\% of the simulations, respectively. In addition, we find that the other three-body angles librate in a greater fraction of the simulations---$\phi_1$ librates in 42.1\% of the simulations (compared to 28.9\%) and $\phi_2$ librates in 83.8\% of our simulations (compared to 72.6\%)---suggesting that the libration of one angle increases the probability of other angles in the system librating. Overall, this change results in a five-planet chain in 27.7\% of our simulations and a 3:2 chain in 99.2\% of our simulations. We compare these percentages to those from our original simulations in Table~\ref{tab:whyf}, referring to this suite as Suite $M_f$.

\textit{Period of f:} We change the orbital period of K2-138f from $12.7576\pm0.0005$ days to $12.757\pm0.005$ days, the final periods of our resonant simulations. We find that this change has a similar effect to altering the mass: planet f is more likely to participate in the chain either through the libration of $\Theta_{f-e}$ or $\phi_3$ (21.2\% and 27.7\% respectively), \textit{and} the other three-body angles are more likely to librate (39.9\% and 82.4\% for $\phi_1$ and $\phi_2$, respectively). This change in orbital period results in a five-planet chain in 24.0\% of our simulations and a 3:2 chain in 99.8\% of our simulations. We compare these percentages to those from our original simulations in Table~\ref{tab:whyf}, referring to this suite as Suite $P_f$.

\textit{Period of e: } Following our K-S and T-test results (see Section~\ref{sec:constrain}), we change the orbital period of K2-138e from $8.2615\pm0.0002$ days to $8.262\pm0.002$ days. This change of period increases the percentage of simulations with librating $\Theta_{f-e}$ or $\phi_3$ to 22\% and 26.5\%, respectively, more than doubling the values. Our simulations result in a 3:2 resonance chain 98.8\% of the time, and 23.4\% of the simulations result in a five-planet chain. We find that the numbers of simulations with librating $\phi_1$ and $\phi_2$ also increase, to 40.5\% and 79.2\% respectively. We compare these percentages to those from our original simulations in Table~\ref{tab:whyf}, referring to this suite as Suite $P_e$.

\textit{Adding K2-138-g:} We did not originally model the outermost planet K2-138g because its sub-Neptune mass ($4.32^{+5.26}_{-3.03}~M_{\oplus}$) and large orbital period compared to K2-138f ($P_g/P_f=3.29$) suggest that it is dynamically decoupled from the rest of the planets in the system. To explore the validity of this assumption, we model K2-138g with a mass of $4.32^{+5.26}_{-3.03}~M_{\oplus}$ and an orbital period of $41.966\pm0.006$ days \citep[][]{Lopez2019}. We find that the addition of planet g indeed increases the chances of planet f participating in the chain, as $\Theta_{f-e}$ and $\phi_3$ librate in 31.7\% and 33.5\% of our simulations, respectively. Similar to the two previous changes, modeling all six planets increases the probability of $\phi_1$ and $\phi_2$ librating to 38.9\% and 82.2\%. However, we find that this additional planet \textit{decreases} the number of simulations with librating  $\Theta_{c-b}$ and $\Theta_{d-c}$ to 48.9\% and 85.0\% respectively, resulting in the dynamical decoupling of K2-138b in 30.0\% of the simulations. Overall, we find that 98.0\% of our simulations result in a 3:2 resonance chain: 24.4\% in a five-planet chain and 56.7\% in a four-planet chain. We compare these percentages to those from our original simulations in Table~\ref{tab:whyf}, referring to this suite as Suite K2-138g.

\begin{deluxetable*}{cccccccc|ccc|c}
\tablecolumns{12}
\tabletypesize{\footnotesize}
\tablecaption{Percentage of resonant simulations\label{tab:whyf}}
\tablehead{
\colhead{Suite} &
\colhead{$\phi_1$} &
\colhead{$\phi_2$} &
\colhead{$\phi_3$} &
\colhead{$\Theta_{c-b}$} &
\colhead{$\Theta_{d-c}$} &
\colhead{$\Theta_{e-d}$} &
\colhead{$\Theta_{f-e}$} &
\colhead{Five-pl} &
\colhead{Four-pl} &
\colhead{Three-pl} &
\colhead{No chain} 
}
\startdata
Orig. & 28.6 & 72.6 & 11.5 & 64.3 & 95.6 & 98.9 & 9.9 & 11.0 & 68.5 & 19.6 & 0.9 \\
$M_f$ & 42.1 & 83.8 & 33.3 & 61.1 & 87.5 & 98.2 & 28.9 & 27.7 & 60.1 & 11.4 & 0.8 \\
$P_f$ & 39.9 & 82.4 & 27.7 & 63.7 & 89.4 & 98.8 & 21.2 & 24.0 & 62.3 & 13.4 & 0.2 \\
$P_e$ & 40.5 & 79.2 & 26.5 & 58.5 & 90.2 & 97.8 & 22.0 & 23.4 & 58.3 & 17.0 & 1.2 \\
K2-138g & 38.9 & 82.2 & 33.5 & 48.9 & 85.0 & 98.4 & 31.7 & 24.4 & 56.7 & 16.8 & 2.0					
\enddata
\tablecomments{Percentage of simulations where each angle librates ($\phi_i$, $\Theta_{i-j}$);  percentage of simulations with a five-, four-, or three-planet 3:2 resonance chain; and  percentage of simulations without a 3:2 resonance chains. We compare our results from the original suite of 3000 \textit{N}-body simulations (Section~\ref{sec:methods}) and the additional four suites (see Section~\ref{sec:whyf}).
See Table~\ref{tab:res_results} for angle definitions.}
\end{deluxetable*}

Although these four changes significantly increase the percentage of simulations where K2-138f is part of the resonance chain, the resulting percentages are still too low to confirm its participation in the chain. Our suite of simulations with increased mass led to the largest percentage of five-planet chains, yet still resulted in K2-138f being dynamically decoupled from the other planets 62.9\% of the time. It is possible that some combination of these effects, or increased precision in the other planets' orbits and masses, will further improve these values. As it stands, we require more data to verify whether or not K2-138f is part of the resonance chain.

\subsubsection{H2: The resonance was broken}

If the K2-138e and K2-138f are not presently in resonance, then perhaps they were at some time but have since been pushed just wide of the resonance. The gravitational interactions between the two planets and their disk, specifically between a planet and the wake of its companion, can reverse convergent migration,  increasing the period ratio between the two planets beyond the resonance width \citep[e.g.][]{Baruteau2013}. Turbulence in the disk can also prevent planets from staying in resonance, sometimes destabilizing the system altogether \citep[][]{Adams2008, Rein2009, Huhn2021}. After the gas disk dispersal, the planetesimal disk or rouge planets are also capable of disrupting the resonant state of planet pairs, whereas less massive planets---such as K2-138f with its mass of $M_p=1.6^{2.1}_{1.2}M_{\oplus}$---are more readily disrupted  \citep[][]{Quillen2013,Chatterjee2015,Raymond2021}.

The loss of energy through tidal dissipation is also an effective means to avoiding or disrupting resonance \citep{Lithwick2012,Lee2013, Delisle2014}, although the result of the system depends on the balance of dissipation in both planets \citep{Delisle2014b}. Tidal dissipation, when combined with secular interactions, can also cause migration of only the innermost planets, leading to divergence; this affect becomes even more dramatic when coupled with a single giant planet on a longer orbital period, although this oftentimes leads to instability \citep{Hansen2015}.

Lastly, it is possible that systems that lock into resonance quickly destabilize after the dispersal of the gas disk without any additional forces \citep[e.g., ][]{Izidoro2017,Izidoro2021}. If the planets' eccentricities are originally damped and the librations are overstable, then a planet pair can readily escape resonance  \citep{Goldreich2014,Delisle2014,Delisle2015}. The ultimate fate of a resonant system's stability is a function of the planet masses, the spacing between the planets, and the number of resonant planets \citep[][]{Matsumoto2012,Deck2015,Pichierri2020}, and might also depend on whether the resonance was formed through inward or outward migration \citep{Lee2009}.

Although there are numerous ways for resonances to be disrupted, studies of resonance \textit{chains} show that any perturbations that are disruptive enough to break a resonance typically lead to chaotic evolution and instability, resulting in a final system architecture that is very different from what we observe \citep{Esteves2020,Huhn2021,Raymond2021}. We therefore find it unlikely that K2-138f could have been removed from the resonance chain.

\subsubsection{H3: K2-138f was never in resonance}
Although convergent migration will generally trap planets into resonance, migration in the absence of effective eccentricity damping can complicate matters. Eccentricities larger than $\sim0.01$---which are comparable to typical eccentricities measured in Kepler planets \citep[e.g.][]{Hadden2014, vaneylen2019}---can make resonance capture more difficult and sometimes impossible for super-Earths \citep{Batygin2015,Pan2017}. \citet{Pan2017} also find that planet pairs that avoid resonance capture are more likely to migrate past and away from each other than they are to collide, leading to a pile-up of planet pairs wide of resonance instead of planet pairs that simply go unstable; however, \citet{Huhn2021} find that such a crossing in resonance chains often leads to rapid eccentricity excitation, which in turn breaks the other resonances in the system.


\subsection{Forming the resonance chain}\label{sec:migration}

While K2-138's resonance chain is interesting, we now question how the resonances formed. Although long-scale migration --- forming the planets more widely spaced and further from their star than observed --- often results in resonance pairs \citep[e.g.,  ][]{snellgrove2001, papaloizou2005, rein2015} and resonance chains \citep{cossou2013} and is often used as the explanation of such chains \citep[e.g., Kepler-223,][]{Mills2017}, resonance chains can potentially form in situ\footnote{Here, we use the term ``in situ'' to distinguish the evolution after the giant impact phase from long-scale migration \citep[e.g., ][]{Lee2002} and to align with other works exploring the in situ formation of super-Earths \citep[e.g., ][]{Dawson2015}. The local growth of these planets does not require that the planetesimals that form them be local, as they could have also accumulated by radial drift.  }, with only small changes to their semi-major axes. \citet{macdonald2018three} explored three different pathways for forming resonance chains, including long-scale migration and two pathways consistent with in situ formation: short-scale migration (where the planets form outside of resonance and small shifts to their orbital periods lock them into resonance) and eccentricity damping. In addition, \citet{Morrison2020} find that close-in super-Earths and mini-Neptunes, such as those in K2-138, can lock into resonance chains due to dissipation from a depleted gas disk and maintain resonance once the gas disk is fully dissipated. Since all currently confirmed resonance chains are consistent with both in situ formation and long-scale migration \citep{macdonald2018three}, we test all three different pathways for forming the resonance chain of K2-138.

Following the methods of \citet{macdonald2018three}, we simulate the formation of this resonance chain via long-scale migration, short-scale migration, and eccentricity damping only. Here, long-scale migration \citep[e.g., ][]{Mills2017} assumes that the planets form far from their star and migrate inwards until they reach their currently observed semi-major axes; short-scale migration \citep[e.g., ][]{MacDonald_2016} assumes that the planets do not undergo significant changes to their semi-major axes during or after the giant impact phase; and eccentricity damping only \citep[e.g., ][]{Dong2016} assumes constant angular momentum. For these three pathways, we apply an inward migration force and/or eccentricity damping forces on timescales $\tau_a\sim10^4$--$10^6$ and $\tau_e\sim10^3$--$10^5$ years, respectively, following the prescription in \citet{Papaloizou2000}. We draw these timescales from independent log-uniform distributions \citep{macdonald2021}. We apply damping forces to only the outer planet\footnote{We do not know the migration rates for the planets since they depend on the conditions of the disk, so we implicitly assume that the migration of the inner planets is on a much longer timescale than that of the outer planet.} for long-scale and short-scale migration, and damp the eccentricities of all planets in the eccentricity damping only simulations. We start all simulations with all planets wide of their observed commensurability and with no librating resonance angles. For each formation pathway, we run 500 simulations, damping the semi-major axes and eccentricities where applicable using the \texttt{modify\_orbits\_forces} routine in the REBOUNDx library. We use the stellar and planetary properties as defined in Table~\ref{tab:sim_parameters}.

\begin{figure*}[ht!]
    \centering
    \includegraphics[width=0.49\textwidth]{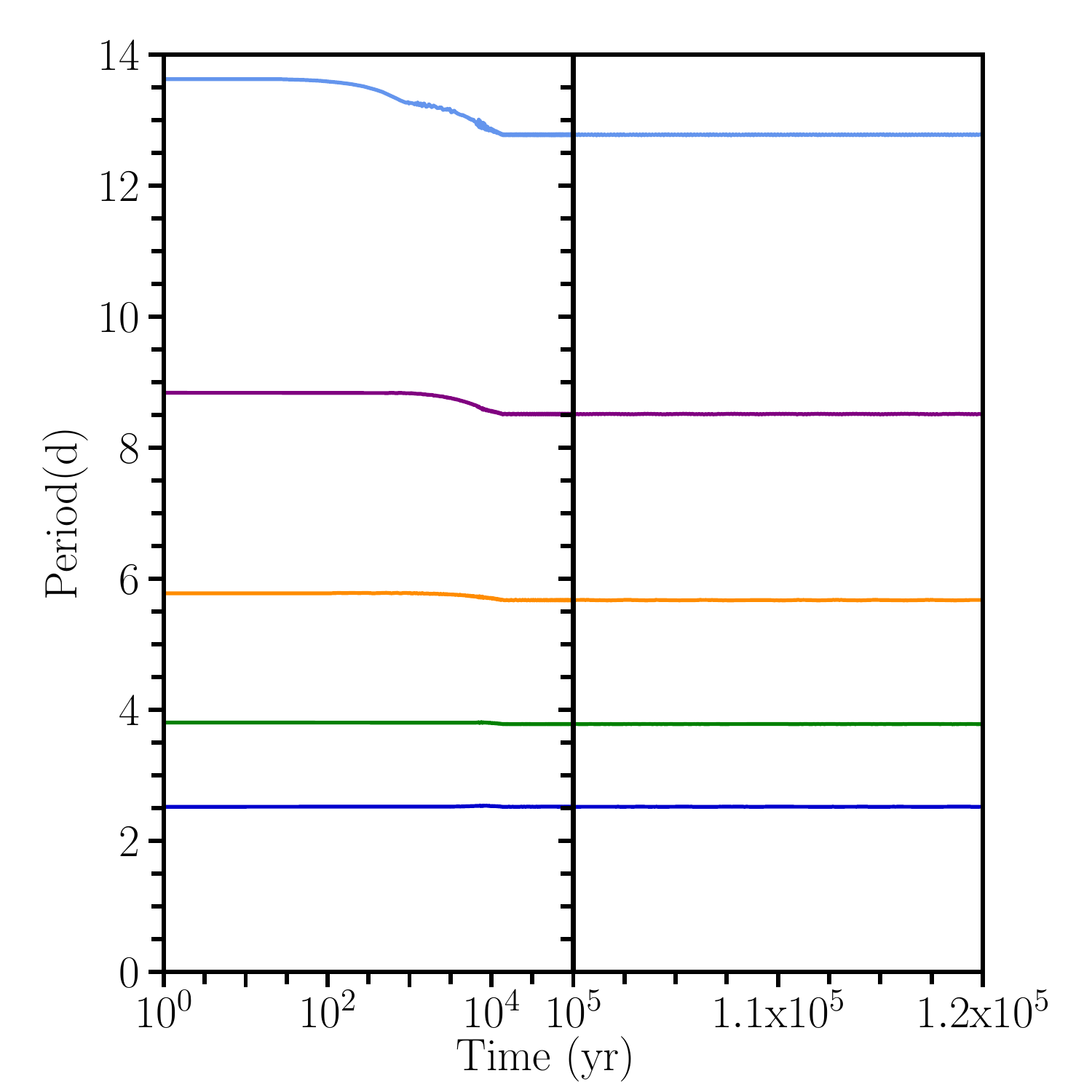}
    \includegraphics[width=0.49\textwidth]{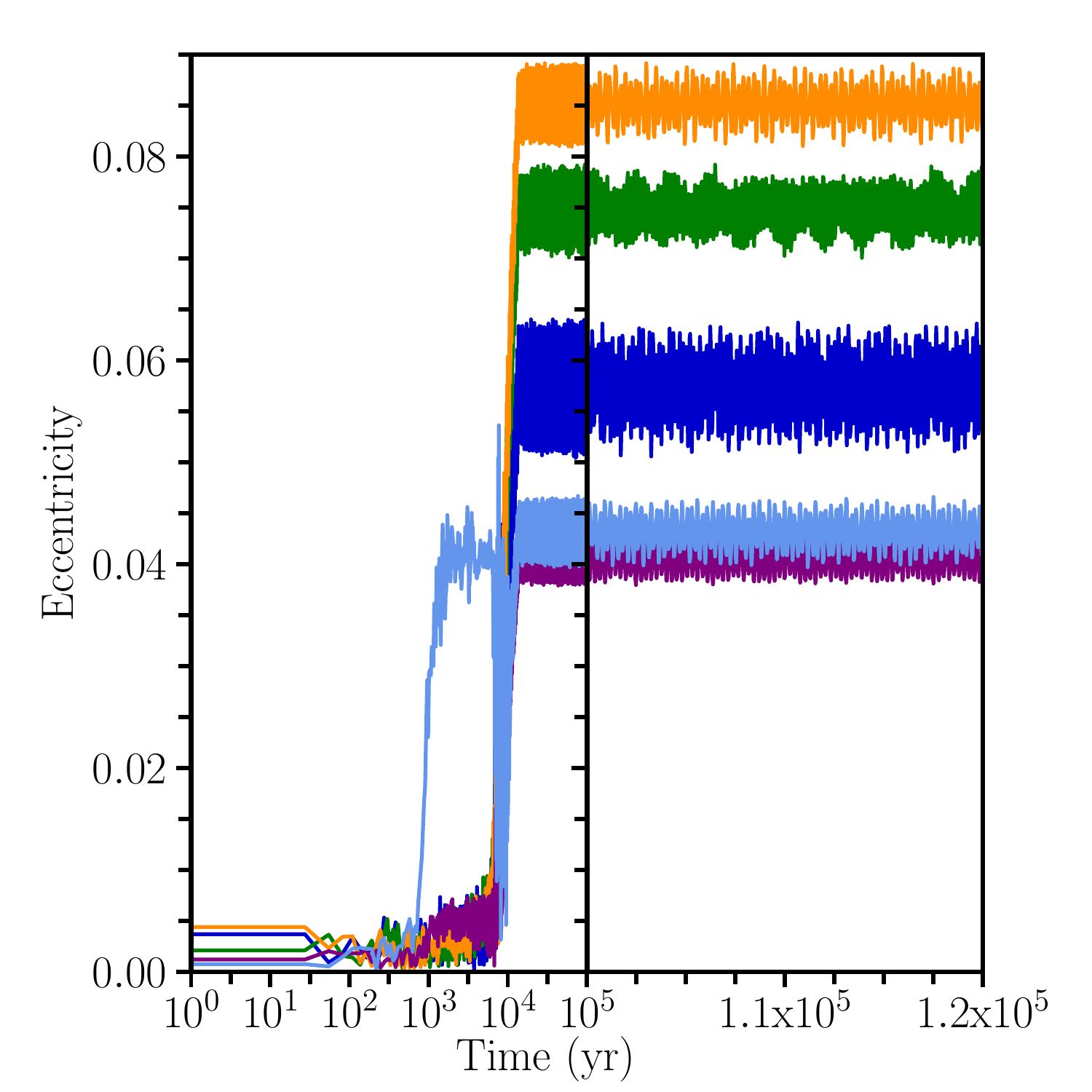}
    \includegraphics[width=0.49\textwidth]{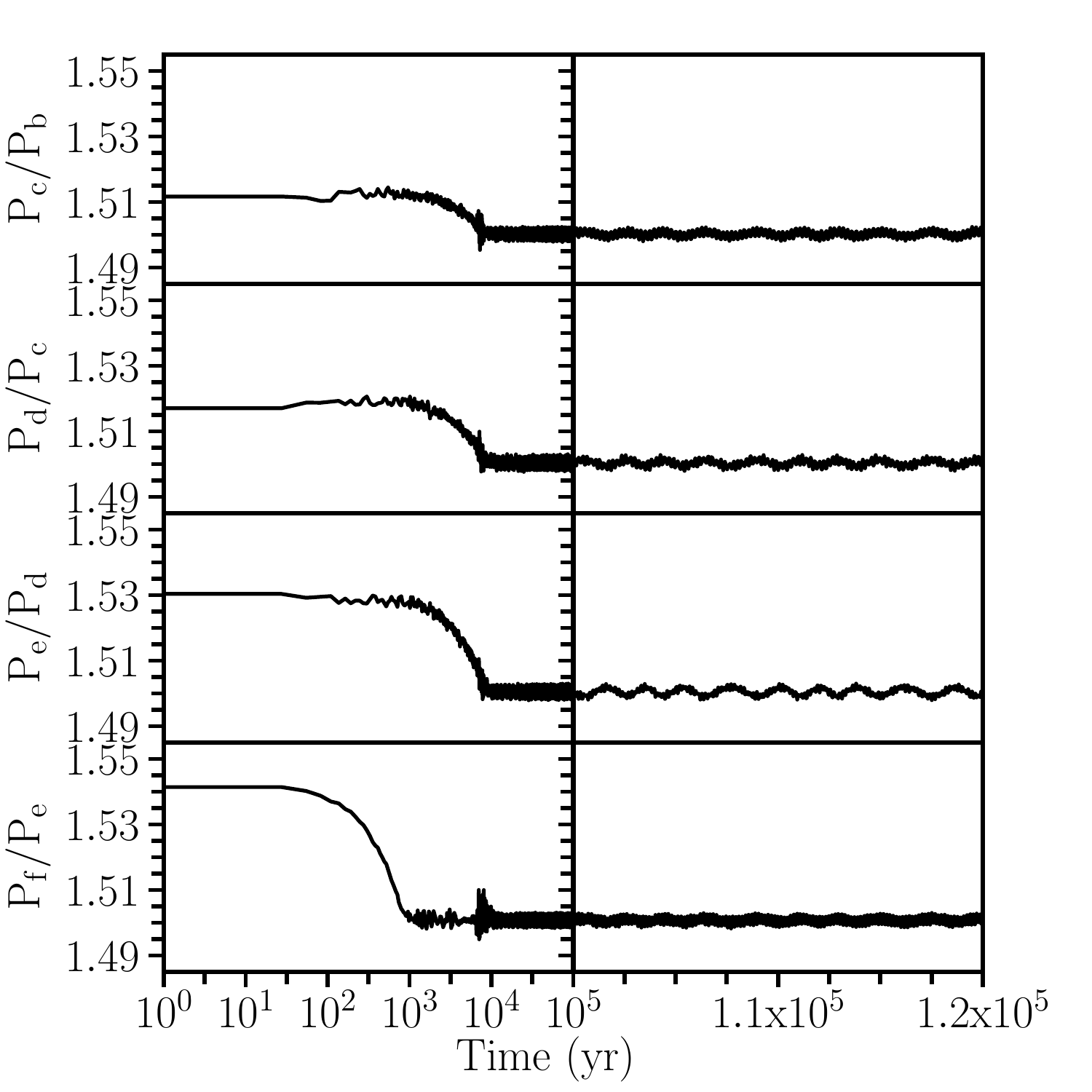}
    \includegraphics[width=0.49\textwidth]{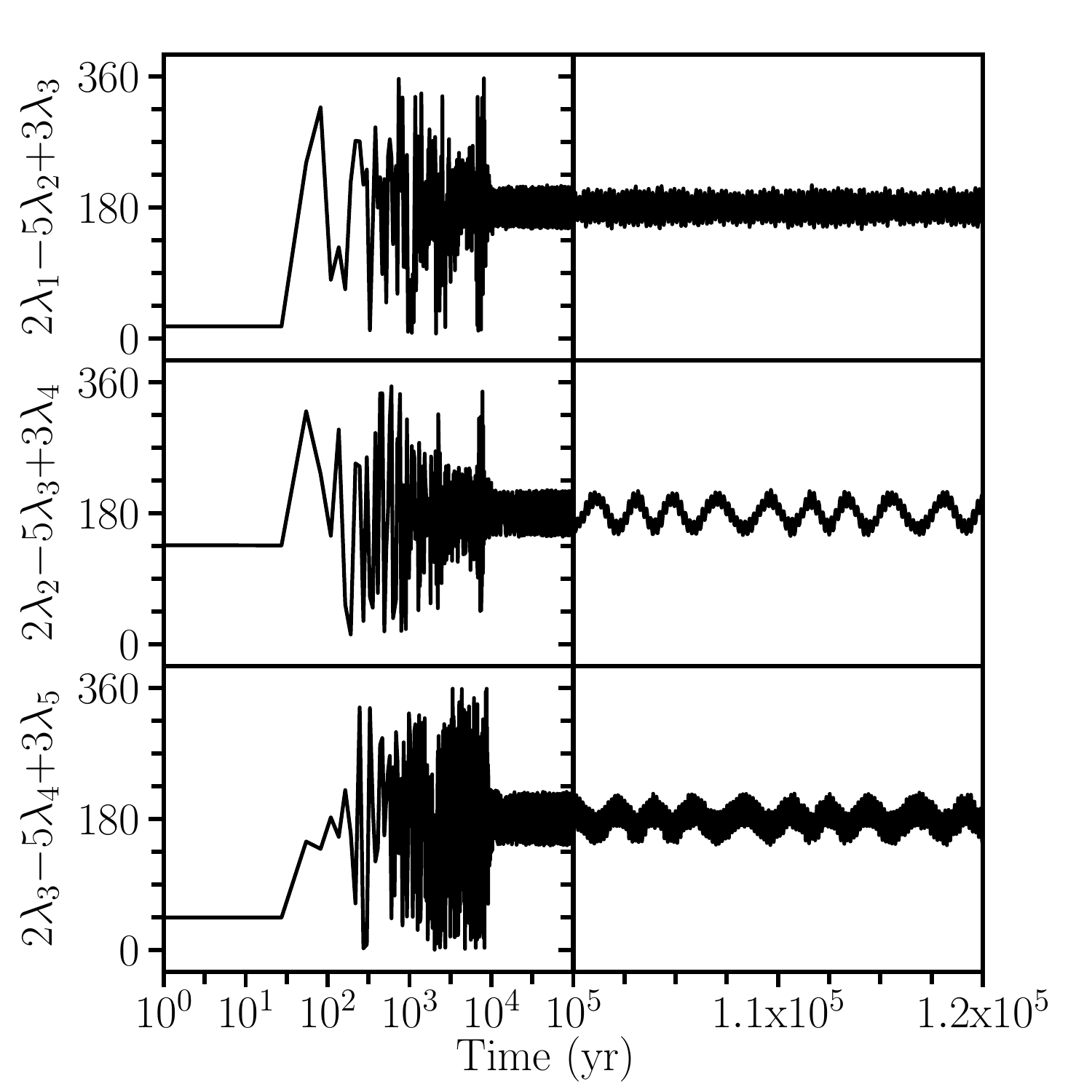}
    \caption{A five planet resonance chain of K2-138 is consistent with in situ formation. Here, we show the evolution of one of our short-scale migration simulations. We start the planets out of resonance and then apply a small damping to the semi-major axis of the outer planet with a timescale of $\tau_a = 4.6 \times 10^4$ years. We force the planets to undergo this small migration for $1.5 \times 10^4$ years, before stopping the migration and allowing the system to evolve for an additional $10^5$ years to confirm long-term stability and resonance. We show the evolution of the orbital periods, eccentricities, period ratios, and three-body resonance angles.}
    \label{fig:k2_migration}
\end{figure*}

We find that we are able to form this resonance chain via all three pathways described above, as each suite of simulations contains numerous sets of initial conditions that lead to fully librating resonance chains. We summarize the resulting centers and amplitudes for the three-body resonance angles in Table~\ref{tab:mig_results}. Since both short-scale migration and eccentricity damping are consistent with in situ formation, we find that the resonance chain of K2-138 could have formed in situ. We show an example simulation from the short-scale migration suite in Figure~\ref{fig:k2_migration}. We caution against using the values in Table~\ref{tab:mig_results} to draw any stronger conclusions, as this study is only able to say what is possible and not what is \textit{more likely.}

\begin{deluxetable}{lcll}
\tablecolumns{4}
\tabletypesize{\footnotesize}
\tablecaption{Resonance Angles from Chain Formation Simulations \label{tab:mig_results}}
\tablehead{
\colhead{Angle} &
\colhead{\% Res} &
\colhead{Center [$\degree$]} &
\colhead{Amplitude [$\degree$]}
}
\startdata
\multicolumn{4}{c}{Short-scale migration, 415/500 stable, 25\% in 5pl resonance}\\
$\phi_1 = 2\lambda_b-5\lambda_c+3\lambda_d$ & 36.63	&	179.9	$^{+	2.7	}_{	2.4	}$ &	45.0	$^{+	26.2	}_{	-20.2	}$ \\
$\phi_2 = 2\lambda_c-5\lambda_d+3\lambda_e$ &  51.81	&	180.0	$^{+	3.8	}_{	3.4	}$ &	35.9	$^{+	23.1	}_{	-13.8	}$ \\
$\phi_3 = 2\lambda_d-5\lambda_e+3\lambda_f$ &  17.83	&	180.4	$^{+	2.4	}_{	1.7	}$ &	33.8	$^{+	24.4	}_{	-11.7	}$ \\
\hline
\multicolumn{4}{c}{Eccentricity Damping, 497/500 stable, 10\% in 5pl resonance}\\
$\phi_1 = 2\lambda_b-5\lambda_c+3\lambda_d$ & 	14.3	&	179.9	$^{+	1.8	}_{	2.7	}$ &	46.8	$^{+	20.8	}_{	-26.7	}$ \\
$\phi_2 = 2\lambda_c-5\lambda_d+3\lambda_e$ &   32.4	&	180.0	$^{+	2.9	}_{	2.0	}$ &	39.0	$^{+	35.3	}_{	-32.8	}$ \\
$\phi_3 = 2\lambda_d-5\lambda_e+3\lambda_f$ &  10.3	&	180.0	$^{+	1.3	}_{	2.2	}$ &	41.6	$^{+	34.8	}_{	-30.8	}$ \\
\hline
\multicolumn{4}{c}{Long-scale migration, 316/500 stable, 53\% 5pl resonance}\\
$\phi_1 = 2\lambda_b-5\lambda_c+3\lambda_d$ & 58.9	&	179.8	$^{+	1.6	}_{	0.6	}$ &	18.3	$^{+	40.1	}_{	-14.2	}$ \\
$\phi_2 = 2\lambda_c-5\lambda_d+3\lambda_e$ &    64.6	&	179.7	$^{+	0.9	}_{	0.9	}$ &	16.1	$^{+	24.5	}_{	-12.6	}$ \\
$\phi_3 = 2\lambda_d-5\lambda_e+3\lambda_f$ & 57.0	&	177.9	$^{+	2.4	}_{	5.5	}$ &	9.8	$^{+	19.6	}_{	-8.3	}$
\enddata
\tablecomments{For each three-body angle $\phi_i$, we include the percentage of the stable simulations where the angle librated and characterize the angle by the center and amplitude of its libration. We report the median and the 68.3\% confidence interval. For each formation mechanism (short-scale migration, eccentricity damping, or long-scale migration), we also report the number of stable simulations and the percentage of those that resulted in a five planet resonance chain. }
\end{deluxetable}

\subsection{Compositions}\label{sec:composition}

\begin{figure}[ht!]
    \centering
    \includegraphics[width=0.48\textwidth]{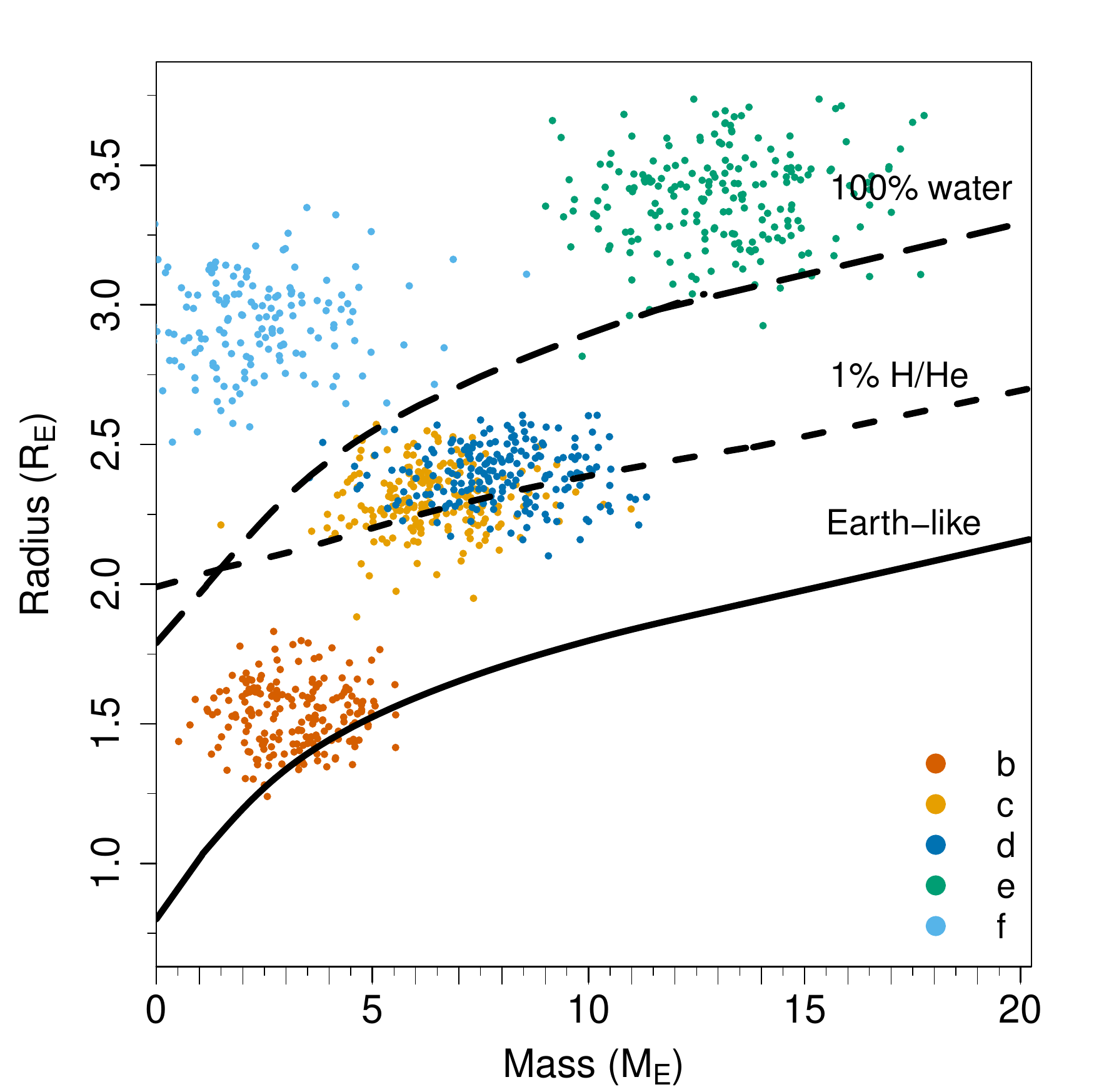}
    \caption{Mass-radius diagram with the radius and mass estimates of K2-138's planets. These estimates were pulled from normal distributions centered on the values from \citet{lopez2019exoplanet}, with standard deviations equal to the uncertainties. We overplot composition curves of Earth-like (solid), 1\% H/He atmosphere (short dash), and 100\% water (long dash). We find that K2-138b is consistent with an Earth-like and terrestrial compositions, but that planets c-f require at least 1\% H/He envelopes to satisfy their densities.}
    \label{fig:k2_MR}
\end{figure}

The compositions of planets can also be used to constrain the formation and evolution of a system. As a first-order approximation of the compositions of K2-138's planets, we first explore the planet's bulk densities, comparing them to Earth-like and less dense compositions. We draw 1000 pairs of mass and radius estimates for the confirmed planets from normal distributions of the parameters in \citet{Lopez2019}, while obeying the 99\% credibility intervals of density. In Figure~\ref{fig:k2_MR}, we plot 200 of these samples with mass-radius curves for compositions of pure water, Earth-like, and 1\% H/He envelopes. We see that K2-138b is consistent with a terrestrial composition while K2-138c, K2-138d, and K2-138e are less dense and require large volatile layers. K2-138f also has a low density, below 2.068 $\mathrm{g\,cm^{-3}}$, and most likely requires the largest atmosphere envelope.

We model the interiors of the four inner planets (b-e) using the planet structure code MAGRATHEA\footnote{We used a preliminary version; the final version will be available at https://github.com/Huang-CL/Magrathea.} (\citealt{Magrathea}), which calculates the pressure, density, temperature, and radius of a spherically symmetric planet with defined mass in each differentiated layer in 1-D using the fourth order Runge-Kutta method. We assume a surface pressure of one bar and use MAGRATHEA's default model. The default model uses equations of state for solid (hexagonal-close-packed\footnote{Under the pressure and temperature of Earth's inner core, the iron most likely takes this structure \citep{vovcadlo1999}}) and liquid iron in the core, bridgmanite and post-perovskite silicate in the mantle, and water, ice-VII, ice-VII', and ice-X in the hydrosphere \citep{Oganov04, Sakai16, Dorogokupets17, Smith18, Grande19}. We model the atmosphere as an ideal gas with a mean molecular weight of 3 $\mathrm{g~cm^{-3}}$, similar to a hydrogen-helium mixture.

To mitigate the degeneracy between water mass fraction and atmospheric mass fraction with core mass fraction, we separate our analysis of each planet into two suites of 1000 models: one where we explore the water mass fraction and one where we explore the atmospheric mass fraction. For our water analysis, we limit the hydrosphere to liquid and ice phases, although the planets' equilibrium temperatures would suggest a vapor layer. We assume an isentropic temperature profile with a surface temperature of 300~K \citep[e.g., ][]{hakim2018}.  For our atmosphere analysis, we assume an ideal gas using an isentropic temperature profile and set the surface temperatures equal to the equilibrium temperatures derived from \citet{Lopez2019}. A real gas would require less atmosphere mass, while a less steep temperature profile would require more. It is important to note that models such as ours are inherently limited for large, hot planets with interior conditions beyond our experimentally-determined equations of state, but although limited, our models are efficient at exploring the range of possible interior solutions.

For each of the 1000 samples of mass and radius, we use a secant method and vary the mass percentage in each layer until the simulated radius matches the sample to 0.01\%. We calculate the water mass fraction (WMF) uniformly across the range of core mass fractions while fixing the core-to-mantle mass ratio. For the atmosphere mass fraction (AMF), we fix the core-to-mantle mass ratio to 1:2 similar to Earth and calculate the AMF uniformly across WMF.

\begin{figure}[ht!]
\plotone{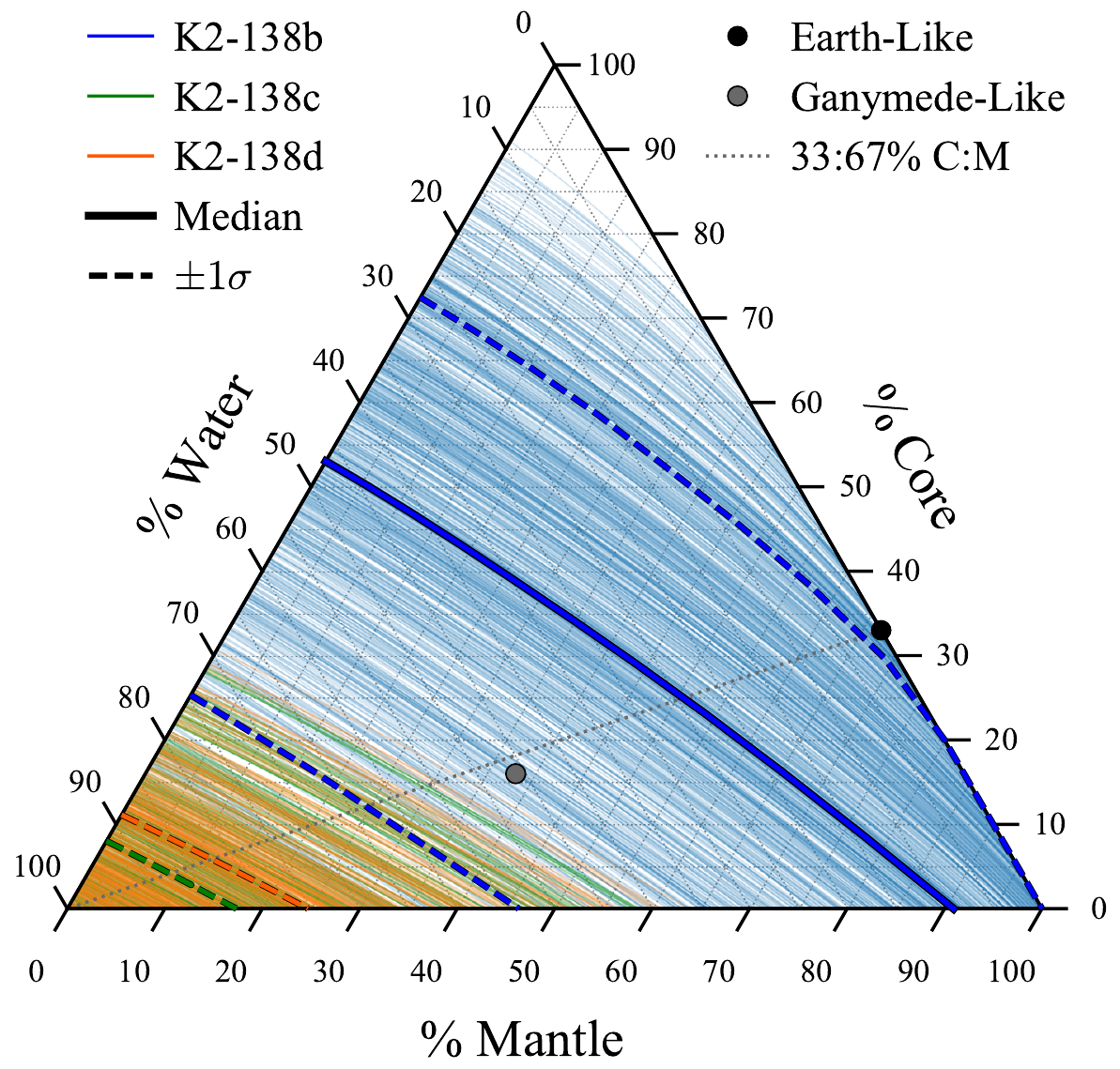}
\caption{Ternary diagram showing solutions from MAGRATHEA to 1000 samples of each planet's observed mass and radius. Here the axes are the percentage of mass in a core, mantle, and hydrosphere. Thin lines show the WMF needed to match the observed radius across core-to-mantle mass ratios. The thick, solid lines correspond to the median WMF while dashed lines are the 1$\sigma$ bounds. The grey dashed line shows a constant Earth-like core-to-mantle ratio. K2-138b (\textit{blue}) is the only planet where all 1000 samples have non-atmosphere solutions. The median samples of K2-138c (\textit{green}) and d (\textit{orange}) require an atmosphere and so we include only the lower 1$\sigma$ bound of WMF. K2-138e requires an atmosphere to satisfy its density and is therefore not included. We use python-ternary by \citet{ternaryplot}.
\label{fig:ternary}}
\end{figure}

We show the resulting WMF in Figure~\ref{fig:ternary}. We find that K2-138b most likely has a WMF between 9.0 and 47\%, depending on the core-to-mantle mass ratio, and an estimated water fraction of 24.3$^{+39.0}_{-22.0}$\% with an Earth-like core-to-mantle mass ratio. 32\% of the samples result in hydrosphere-free solutions, meaning the planet could be composed of only core and mantle and no water. With Earth-like core-to-mantle mass ratios, only 18.3\% of the K2-138c samples and 25.5\% of the K2-138d samples have three-layer solutions with less than 90\% WMF, suggesting appreciable atmospheres. For reference, models of Neptune suggest at least 80\% mass in a water-dominated fluid layer \citep{Scheibe19}.

\begin{figure}[ht!]
\includegraphics[width=0.49\textwidth]{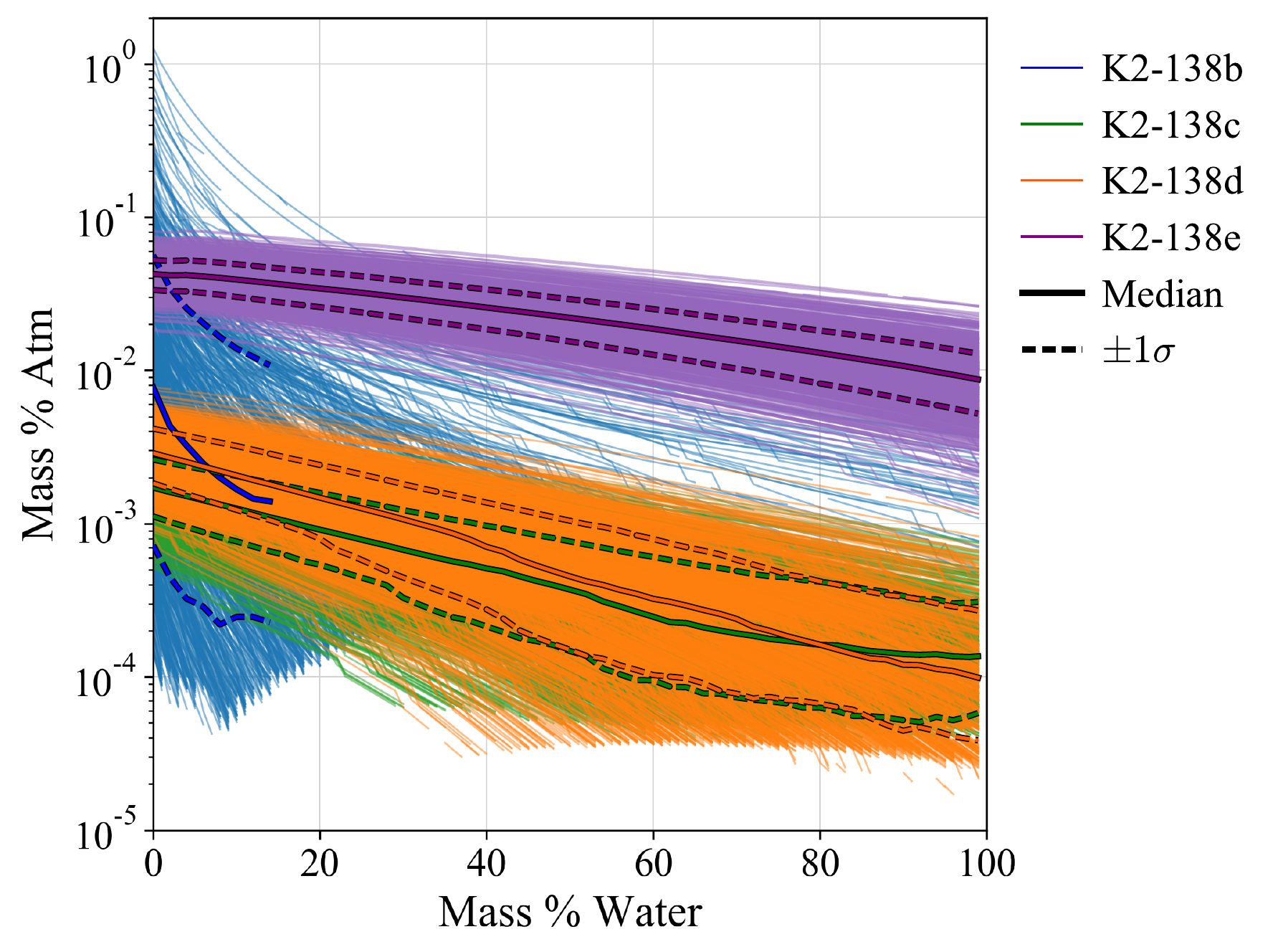}
\caption{The water and atmosphere mass fractions needed to match samples of mass and radius for each planet. We fix the core-to-mantle mass ratio to 1:2, similar to Earth. Thin background lines are solutions to each sample, and thick lines are the mean (solid) and 1$\sigma$ bounds (dashed) of samples with solutions at the given water mass. We use an isentropic temperature profile with the surface temperature set to the equilibrium temperature from \citet{Lopez2019}. We only include the statistics for K2-138b until more than 50\% of the samples result in solutions that require less than 10$^{-4}$\% atmospheric mass. 19\% of K2-138b samples are too dense to require an atmospheric mass of more than 10$^{-4}$\% and are not shown.
\label{fig:atm}}
\end{figure}

In Figure~\ref{fig:atm}, we show the AMF needed to match samples of the planets' masses and radii across WMF. Across all possible water mass fractions, we predict AMF of the four planets of: 1.7$^{+9.3}_{-0.9}\times10^{-3}$\%, 5.3$^{+7.9}_{-3.8}\times10^{-4}$\%, 7.0$^{+13}_{-5.6}\times10^{-4}$\%, 0.022$^{+0.016}_{-0.011}$\%. K2-138b's mass and radius could be explained with a hot, inflated H/He atmosphere layer, but many solutions require atmospheric masses of less than 10$^{-6}$ M$_\oplus$, 1.25\% of Venus's atmospheric mass. With WMF~=~0, K2-138c and K2-138d require an AMF of over 0.001\%. The pressure and temperature under the atmosphere of K2-138c with 50\% WMF is around 10~bar and 4000~K which, in our model, creates a small layer of liquid water before transitioning to high-pressure ices in the model. This temperature and pressure suggest that the water would be gaseous, but understanding this boundary requires a model that couples the atmosphere and interior. K2-138e requires an AMF around 15-30 times more than K2-138c and d at zero WMF.

Aside from the similar inferred compositions of K2-138c and K2-138d, the possible compositions of the planets of K2-138 have little overlap. Their densities decrease and inferred volatile content increases with orbital period. Other systems with resonances, such as TRAPPIST-1, and other compact multi-planet systems have inter-planetary similarities in their sizes and masses and therefore in their inferred compositions \citep{weiss2018,millholland2021,Agol21}. This similar sizing within systems, especially when paired with the regular orbital spacing these systems also exhibit, could be telling of the formation history and/or the subsequent dynamical evolution \citep[e.g., ][]{adams2019,macdonald2020,mishra2021}, and K2-138's lack of intra-system uniformity could be just as telling.

\section{Conclusion}\label{sec:conclusion}

The K-dwarf K2-138 hosts six confirmed planets, all with period ratios of adjacent planets near the 3:2 commensurability. We run numerical \textit{N}-body simulations using REBOUND \citep{rebound}, drawing the planetary parameters from normal distributions centered on the results from \citet{lopez2019exoplanet}, and modeling the system for 8Myr. We analyze our resulting simulations, finding that nearly all of our simulations (99.2\%) result in a chain of 3:2 resonances, although few (11.0\%) result in a five planet chain. We find that K2-138b and K2-138f do not need to be in resonance for the system to be stable, as 87.1\% of our simulations result in K2-138f being dynamically decoupled from the other planets and 21.4\% of the simulations result in K2-138b being dynamically decoupled. We are, however, able to confirm a resonance chain of 4:6:9 between K2-138c, K2-138d, and K2-138e, as 99.0\% of our simulations result in the libration of the angle $\phi_2 = 2\lambda_c-5\lambda_d+3\lambda_e$ and/or the libration of both two-body resonance angles.


Although numerous mechanisms exist for breaking a potential past resonance between K2-138e and K2-138f, resonance breaking within resonance chains usually leads to breaking the other resonances in the systems and oftentimes to instability. We argue that it is then more likely that K2-138f was never in resonance or that it is resonant and we simply have insufficient data to prove the resonance. Additional photometry could tease out the TTVs---which are expected to be on the order of 2--5 minutes but continue to be elusive---that would produce a stronger signal to be fit, further constraining the planets' masses and orbital parameters. The resulting decrease in parameter uncertainties could be sufficient to confirm additional resonances in this system, including the three-body resonance between planets d, e, and f.

We analyze our simulations for links between initial planetary parameters and resulting resonance or stability. We find that all resonance angles aside from $\Theta_{e-d}$ show preference for specific masses and orbital periods. K2-138b, K2-138c, and K2-138f require slightly larger masses than those estimated by \citet{Lopez2019} for K2-138b and K2-138f to be in resonance, and K2-138e requires a slightly larger period for K2-138f to part of the chain. Through additional \textit{N}-body simulations, we find that we can increase the number of simulations where K2-138f is part of the resonance chain by increasing its mass, by altering its orbital period, by increasing the orbital period of K2-138e, and by modeling K2-138g; however, none of these changes alone is sufficient in confirming that K2-138f is in resonance.

Resonance chains are often seen as the hallmark of disk migration, but previous studies have found that such dynamical configurations can also arise from in situ formation \citep{MacDonald_2016,Dong2016,macdonald2018three}. We therefore explore whether a five-planet chain of 3:2 resonances could have formed in situ as well as through migration. We find that K2-138 and its dynamics are consistent with in situ formation, but could have also been formed through long scale migration. Additional data of the system could potentially constrain the formation history, depending on the resulting resonance centers and amplitudes.

Using the planet structure code MAGRATHEA, we then explore the potential compositions of the planets along the uncertainties in their masses and radii. We find that K2-138b is consistent with a terrestrial composition and that any atmosphere would be less than half the mass of Venus's atmosphere. K2-138c and K2-138d have similar compositions; both planets require a minimum of $\sim$80\% of their mass to be water, or an atmosphere, with over 0.01\% of their mass as atmospheres. The bulk density of K2-138e is inconsistent with a non-atmospheric model; without a hydrosphere, we estimate an atmospheric mass fraction of 0.5\%, or 0.065$M_{\oplus}$.

The confirmation of additional resonance chains could help us constrain the formation history of the individual systems and identify indicators of formation history in other systems. With the new planets discovered by TESS and those to be discovered in the near future, we will soon have a sufficient number of resonance chain systems to leave the area of small-number statistics and begin a full scale study of the formation history of exoplanetary systems.

We thank the referee for their constructive review that significantly improved this work and Kevin Hardegree-Ullman for sharing his detrended lightcurve. DR would like to thank Chenliang Huang and Jason H. Steffen for their advice and discussion. MGM acknowledges that this material is based upon work supported by the National Science Foundation Graduate Research Fellowship Program under Grant No. DGE1255832. Any opinions, findings, and conclusions or recommendations expressed in this material are those of the author and do not necessarily reflect the views of the National Science Foundation. The authors acknowledge use of the ELSA high performance computing cluster at The College of New Jersey for conducting the research reported in this paper. This cluster is funded in part by the National Science Foundation under grant numbers OAC-1826915 and OAC-1828163. 

\newpage
\appendix

A planetary mass estimate requires radial velocity follow-up or for the planets to be gravitationally perturbing each other's orbits enough that we can detect significant variations in the transit times, or TTVs. Using HARPS spectra in combination with the Kepler photometry, \citet{lopez2019exoplanet} constrained the masses of the planets of K2-138, and both \citet{christiansen2018k2} and \citet{lopez2019exoplanet} estimate TTVs on the order of 2-5 minutes. However, the TTVs have so far been elusive and might require higher cadence observations.

One step beyond fitting the system's TTVs would be to forward model and fit the lightcurve itself, in a manner that is both contained and self-consistent. \citet{Mills2017} fit the lightcurve of Kepler-444 using a photodynamic model, constraining two of the planet masses and the orbital elements for all of the planets. Such a method can be employed to other systems. By directly forward modeling and fitting the lightcurve of this system, following \citet{Mills2017} and using PhoDymm (Ragozzine et al., in prep.), we aim to determine the masses and orbits for all five planets hosted by K2-138. We follow the methods outlined in \citet{Mills2017} which we summarize below.

We integrate the Newtonian equations of motion for K2-138 and its five confirmed planets. We generate a synthetic lightcurve from a limb-darkened lightcurve model to compare to the K2 photometry reduced by \citet{Hardegree2021} and perform Bayesian parameter inference using differential equation Markov Chain Monte Carlo \citep[DEMCMC, ][]{Ter2006}. We fit the orbital period, the mid-transit time, the eccentricity, the argument of periapse, the sky-plane inclination, the radius, and the mass of each planet. In addition, we fit the star's mass and radius, the two limb-darkening coefficients, and the amount of dilution from other nearby stars. We employ Gaussian priors on the stellar mass and radius based on values from \citet{christiansen2018k2} ($M = 0.93\pm 0.06 M_{\odot}$, $R=0.86\pm 0.08 R_{\odot}$). We also fix $\Omega=0$ for all planets, given that the system has small mutual inclinations and employ flat priors on all other parameters.

We run a 96-chain DEMCMC for 450,000 generations, recording every 1000th generation and removing a burn-in of 20,000 generations. We include the median and 68.3\% confidence intervals from the phoyodynamic model in Table~\ref{tab:phodymm}. 

Unfortunately, we are unable to constrain the planet masses to any useful precision. We suspect this is due to the low signal of the gravitational perturbations between planets that also drives the missing TTVs. We also find that the planetary radii are greatly overestimated but with small precision --- for example, our estimate of the radius of K2-138c is $R_c~=~8.1\pm0.2R_{\oplus}$ compared to $2.299^{+0.12}_{-0.087}R_{\oplus}$ from \citet{Lopez2019} --- suggesting that the values are overfit. Such a study should be repeated once higher precision or higher cadence data are available.

\begin{deluxetable*}{llllll}[ht!]
\renewcommand{\arraystretch}{0.75} 
\tablecolumns{6}
\tablewidth{0pt}
\tabletypesize{\small}
\tablecaption{ Photodynamic fitting results from PhoDyMM \label{tab:phodymm}}
\tablehead{
\colhead{} &
\colhead{K2-138b} & 
\colhead{K2-138c} & 
\colhead{K2-138d} &
\colhead{K2-138e} &
\colhead{K2-138f}
} 
\startdata
$M_p~(M_{\oplus})$	&	6.3	$^{+	13.5	}_{-	5.0	}$ &	4.3	$^{+	10.9	}_{-	3.3	}$ &	9.0	$^{+	10.3	}_{-	5.9	}$ &	5.0	$^{+	7.0	}_{-	3.8	}$ &	10.6	$^{+	12.2	}_{-	7.4	}$ \\
$R_p~(R_{\oplus})$	&	5.4	$^{+	0.1	}_{-	0.1	}$ &	8.1	$^{+	0.2	}_{-	0.2	}$ &	8.6	$^{+	0.2	}_{-	0.2	}$ &	10.9	$^{+	0.2	}_{-	0.2	}$ &	9.6	$^{+	0.2	}_{-	0.2	}$ \\
$P$ (d)	&	2.353	$^{+	0.0003	}_{-	0.0003	}$ &	3.561	$^{+	0.001	}_{-	0.001	}$ &	5.404	$^{+	0.001	}_{-	0.002	}$ &	8.263	$^{+	0.002	}_{-	0.001	}$ &	12.759	$^{+	0.001	}_{-	0.001	}$ \\
$t_0$ (d)	&	773.317	$^{+	0.003	}_{-	0.003	}$ &	740.309	$^{+	0.009	}_{-	0.012	}$ &	743.162	$^{+	0.009	}_{-	0.005	}$ &	740.642	$^{+	0.004	}_{-	0.006	}$ &	738.700	$^{+	0.003	}_{-	0.003	}$ \\
i ($\degree$)	&	87.02	$^{+	1.41	}_{-	0.93	}$ &	88.22	$^{+	0.85	}_{-	0.63	}$ &	89.48	$^{+	0.37	}_{-	0.57	}$ &	88.83	$^{+	0.33	}_{-	0.29	}$ &	88.88	$^{+	0.20	}_{-	0.19	}$ \\
$e$	&	0.002	$^{+	0.04	}_{-	0.04	}$ &	0.002	$^{+	0.03	}_{-	0.03	}$ &	0.001	$^{+	0.02	}_{-	0.02	}$ &	0.007	$^{+	0.03	}_{-	0.02	}$ &	0.011	$^{+	0.02	}_{-	0.03	}$ \\
$\omega$ ($^{\circ}$)	&	74.0	$^{+	45.2	}_{-	54.0	}$ &	-100.2	$^{+	52.0	}_{-	48.0	}$ &	107.8	$^{+	46.8	}_{-	47.3	}$ &	-112.7	$^{+	49.6	}_{-	42.9	}$ &	27.2	$^{+	52.6	}_{-	45.1	}$
\enddata
\tablecomments{Here, $M_p$ is the planetary mass, $R_p$ is the planetary radius, $P$ is the orbital period, $t_0$ is the transit epoch (BJD$-2457000$),   $i$ is the sky-plane inclination, $e$ is the eccentricity, and $\omega$ is the argument of periapsis.}
\end{deluxetable*}

\newpage

\bibliographystyle{aasjournal}
\bibliography{main}

\end{document}